\documentclass{aastex631}
\usepackage{CJK}

\usepackage{newtxtext,newtxmath}
\usepackage[T1]{fontenc}
\usepackage{ae,aecompl}
\usepackage{graphicx}	
\usepackage{amsmath}	
\usepackage{amssymb}	
\usepackage{xcolor}

\newcommand{\szpsc}{\mbox{SZ~Psc}}

\newcommand{\vsini}{\mbox{$v \sin i$}}

\received{}
\revised{}
\accepted{}
\submitjournal{ApJ}

\shorttitle{Doppler imaging of SZ Psc}
\shortauthors{Xiang et al.}

\begin{document}
\begin{CJK*}{UTF8}{gbsn}
\title[Doppler imaging of \szpsc]{Further study of starspot activity and measurement of differential rotation for SZ Piscium}

\correspondingauthor{Shenghong Gu, Yue Xiang}
\email{shenghonggu@ynao.ac.cn, xy@ynao.ac.cn}

\author{Yue Xiang}
\affiliation{Yunnan Observatories, Chinese Academy of Sciences, Kunming 650216, China}
\affiliation{Key Laboratory for the Structure and Evolution of Celestial Objects, Chinese Academy of Sciences, Kunming 650216, China}
\affiliation{International Centre of Supernovae, Yunnan Key Laboratory, Kunming 650216, China}

\author{Shenghong Gu}
\affiliation{Yunnan Observatories, Chinese Academy of Sciences, Kunming 650216, China}
\affiliation{Key Laboratory for the Structure and Evolution of Celestial Objects, Chinese Academy of Sciences, Kunming 650216, China}
\affiliation{International Centre of Supernovae, Yunnan Key Laboratory, Kunming 650216, China}
\affiliation{School of Astronomy and Space Science, University of Chinese Academy of Sciences, Beijing 101408, China}

\author{A. Collier Cameron}
\affiliation{School of Physics and Astronomy, University of St Andrews, Fife KY16 9SS, UK}

\author{J. R. Barnes}
\affiliation{Department of Physical Sciences, The Open University, Walton Hall, Milton Keynes MK7 6AA, UK}

\author{Dongtao Cao}
\affiliation{Yunnan Observatories, Chinese Academy of Sciences, Kunming 650216, China}
\affiliation{Key Laboratory for the Structure and Evolution of Celestial Objects, Chinese Academy of Sciences, Kunming 650216, China}
\affiliation{International Centre of Supernovae, Yunnan Key Laboratory, Kunming 650216, China}

\begin{abstract}
We present a series of 9 Doppler images of the magnetically active K component of the RS CVn-type binary SZ~Psc, based on the high-resolution spectroscopic data collected from 2014 to 2018. We apply least-squares deconvolution to all spectra to extract the average profiles with high signal-to-noise ratios (SNRs) for Doppler imaging. The surface maps of the K subgiant show starspots widely distributed along latitude and longitude. A prominent, non-axisymmetric polar spot around phase 0 is revealed by all images with sufficient phase coverage, which may be a stable feature on the K component. The starspots evolve in a time scale of one month. We have determined the surface shear rate of the K component from the starspot maps reconstructed 10 days apart in 2017 Nov--Dec, through the cross-correlation method. The surface differential rotation parameters are $\Omega_{eq} = 1.591 \pm 0.002$ rad d$^{-1}$ and $\Delta \Omega = 0.035 \pm 0.003$ rad d$^{-1}$. The absorption lines contributed from the tertiary component are detected in all LSD profiles of SZ~Psc, and we measure the radial velocity of the binary system and the tertiary component to derive an elliptical orbit with a period of $1530 \pm 3$ days and a mass of $0.75 \pm 0.06$ M$\odot$ for the tertiary component.
\end{abstract}


\section{Introduction}
Starspot, which is sunspot analogue, is an observable tracer of the stellar magnetic field on the photosphere. Imaging studies of starspots on different single and binary stars are essential to understand the magnetic dynamo process undertaking beneath the stellar surface \citep{berdyugina2005,strassmeier2009}. With a time-series of high-resolution spectra, Doppler imaging can reconstruct a stellar surface map with much more details than that derived from light curve inversions.

SZ Psc is a very active RS CVn-type binary, consisting with a F8V and a K1IV components, with a period of nearly 4 days \citep{eaton2007}. The cooler component (K1IV) is magnetically active and shows spot and flare activities \citep{xiang2016,karmarkar2023}. The brightness of SZ Psc is varying due to the presence of starspots on its surface \citep{kang2003}. The starspot activtiy of SZ Psc has been widely investigated through the photometric analysis (e.g. \citealt{lanza2001,kang2003,eaton2007}). \citet{kang2003} revealed that the shape of light curve was changing due to the migration and evolution of starspots. The chromosphere of the K subgiant of SZ~Psc is very active, and shows emission in activity spectral indicators, such as Na I D1, D2, H$\alpha$ and Ca II IRT \citep{zhang2008,cao2012,cao2019,cao2020}. \citet{cao2019} detected potential prominence activity on the cooler component of SZ Psc from the time-series of H$\alpha$ profiles.

Differential rotation is one of the most important ingredients of the stellar magnetic dynamo process, and the measurement of surface differential rotation rate is an important goal in recent Doppler imaging studies (e.g. \citealt{perugini2021,finociety2023,kriskovics2013}). The shear rate can be measured from the cross-correlation between the surface patterns of brightness or magnetic field, obtained in close-by epochs \citep{donati1997b}, or as imaging parameters that are directly taken into account in the reconstruction procedure \citep{petit2002}. The surface differential rotation of the single and binary stars are derived by some authors \citep{kovari2017}.

In the previous work, we performed the first Doppler imaging study of SZ~Psc based on the data acquired in 2004 November, 2006 September, November and December \citep{xiang2016}. The surface images show a more complex spot patterns on the surface of the magnetically active K component than those revealed by the light curve modelling. The weak absorption line of the third component was also detected in all observations. In this work, we present further Doppler imaging study on the starspot activity and the first measurement for differential rotation of SZ Psc. We also use the longer data sets obtained during 2014--2018 to further investigate the property of the third component of the system. The description of the spectroscopic observations and data reduction is given in Section 2. We present the Doppler images and the measurement of differential rotation in Section 3. In section 4, we analyse the radial velocity curve of the third component of SZ Psc from the LSD profiles, considering the previous results. At last, we give the summary and discussion on our work in Section 5.

\section{Observations and data reduction}

The high-resolution spectroscopic observations on SZ~Psc were carried out from 2014 to 2018 at 2.16m telescope of the Xinglong Station, National Observatories of China, and 2.4m telescope \citep{fan2015} of the Lijiang station, Yunnan Observatories. The same type fibre-fed high-resolution spectrographs (HRS), equipped with a 4096$\times$4096 CCD detector, were used to collect spectra by both telescopes. The spectral resolution of HRS is $\sim$48000 and the spectral coverage is 3900--9500\AA. These spectroscopic observations are the same as those in \citet{cao2020}, who investigated the chromospheric activity of SZ~Psc and thus concentrated on some stellar activity indicators. We give a summary of the observations on SZ~Psc in Table \ref{tab:log}, which includes the UT date of the beginning and end of each observing run, the number of the observed spectra, the typical exposure time, and the average signal-to-noise ratio (SNR) of the raw spectra. In addition, the spectra of two inactive stars, HR 7690 and HR 7560, were observed at the same time as the templates to mimic the photospheres of F and K components of the system, which are required by the used Doppler imaging method. These spectra were also used by \citet{cao2020} as the templates in the spectral subtraction procedure, where they demonstrated that the inactive photospheric spectra of F and K components can be well synthesized by using these two spectral templates.

\begin{deluxetable}{lccccccc}
\tabletypesize{\scriptsize}
\tablecolumns{4}
\tablewidth{0pt}
\tablecaption{A summary of the observations on SZ Psc.}
 \label{tab:log}
\tablehead{
 \colhead{Start date}&
 \colhead{End date}&
 \colhead{No. of Spectra}&
 \colhead{Exposure}&
 \colhead{Avg. SNR}&
 \colhead{Avg. SNR}&
 \colhead{RV$_{\rm binary}$}&
 \colhead{RV$_{\rm tertiary}$}\\
 \colhead{UT}&
 \colhead{UT}&
 \colhead{}&
 \colhead{s}&
 \colhead{Raw}&
 \colhead{LSD}&
 \colhead{km/s}&
 \colhead{km/s}
}

\startdata
  2014 Oct 05 & 2014 Oct 06 & 18 & 1800 &  63 & 1504 & 15.6 & -12.9\\
  2015 Oct 27 & 2015 Nov 02 & 74 &  900 &  85 & 2027 &  6.7 &  23.0\\
  2016 Nov 13 & 2016 Nov 19 & 10 &  900 & 120 & 2875 &  5.5 &  26.5\\
  2016 Dec 16 & 2016 Dec 19 & 18 & 1800 & 125 & 2975 &  6.3 &  24.9\\
  2017 Nov 01 & 2017 Nov 04 & 20 & 1800 &  47 & 1120 & 10.0 &   7.4\\
  2017 Nov 28 & 2017 Dec 01 & 45 &  900 &  70 & 1674 & 11.1 &   5.1\\
  2017 Nov 05 & 2017 Dec 11 & 67 & 1800 &  61 & 1469 & 11.6 &   3.9\\
  2017 Dec 28 & 2018 Jan 01 & 11 & 1800 & 121 & 2880 & 12.3 &   2.3\\
  2018 Nov 21 & 2018 Nov 25 & 36 & 1800 &  84 & 2009 & 16.9 & -14.4\\
\enddata
\end{deluxetable}

All raw CCD images collected by the spectrographs were reduced by using the IRAF\footnote{IRAF is distributed by the National Optical Astronomy Observatory, which is operated by the Association of Universities for Research in Astronomy (AURA) under cooperative agreement with the National Science Foundation.} package in the standard procedure, which includes image trimming, bias and flat field corrections, scatter light subtraction, cosmic-ray removal, one-dimension spectra extraction and continuum fitting.

A single exposure of an echelle spectrograph can record thousands of stellar photospheric lines. Least-squares deconvolution (LSD) extracts a mean profile from all available atomic lines, and significantly improves the SNR of the stellar line profile \citep{donati1997}. We applied the LSD to all observed spectra. The line list was extracted from the Vienna Atomic Line Database (VALD; \citealt{kupka1999,ryabchikova2015}) with the same atmospheric parameters used in \citet{xiang2016}. With a wider wavelength coverage, the SNR gain is larger than that in \citet{xiang2016}. The spectral lines that fall into the regions of strong telluric lines and active chromospheric lines, such as H$\alpha$, Ca II IRT, etc. were removed from the final list. As a result, 3235 lines were used in each LSD computation. The average SNR of LSD profiles of each observing run is also listed in Table \ref{tab:log}. We also derived the LSD profile of the telluric lines in each spectra to estimate the instrumental drift in wavelength calibration, and then corrected the shift to suppress the error in radial velocity measurement.

\begin{figure}
\centering
\includegraphics[width=0.45\textwidth]{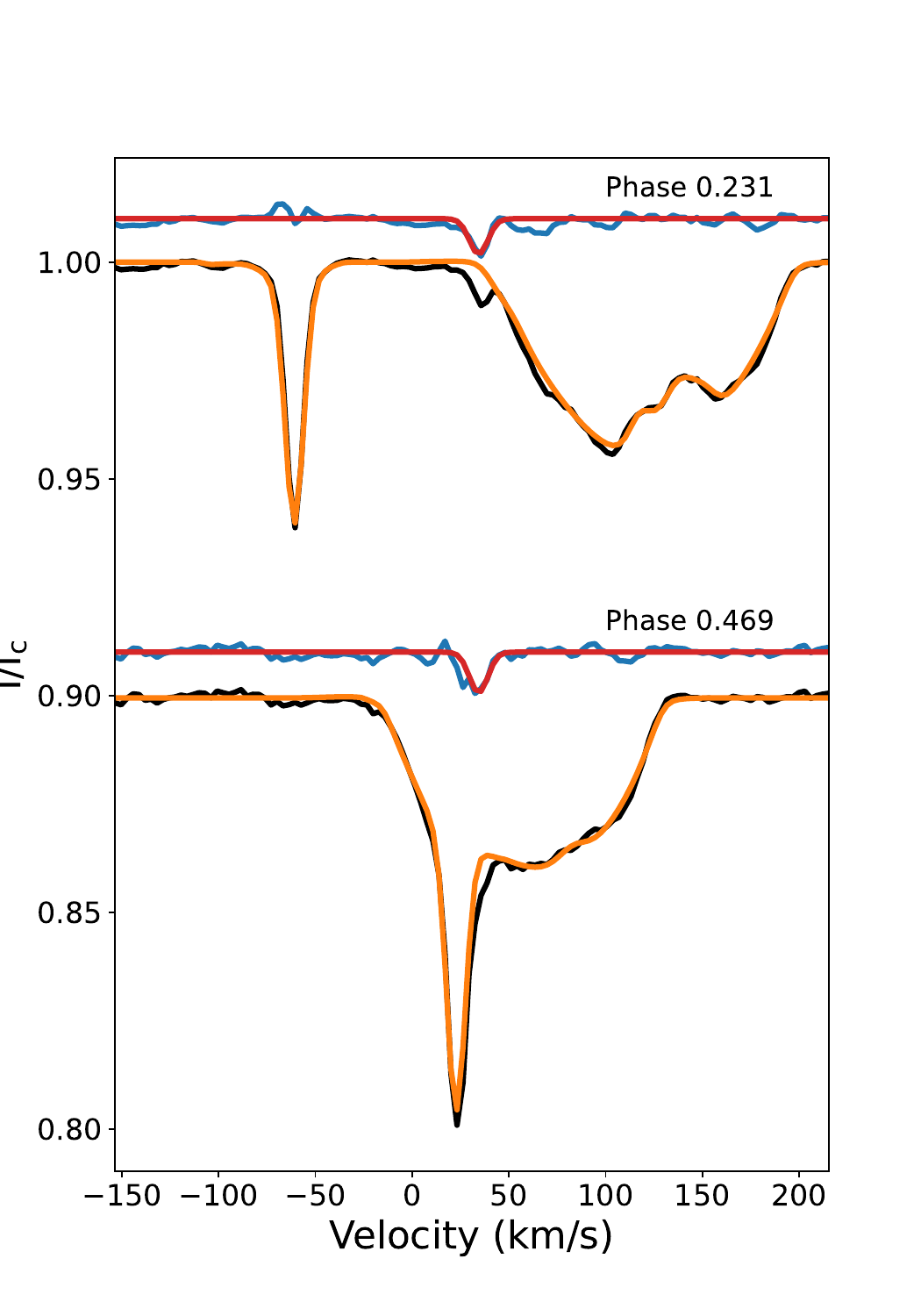}
\caption{The black lines represent the LSD profiles of SZ Psc at two different phases, while the orange lines show the initial fits. The blue and red lines above each LSD profile indicate the residual and the corresponding Gaussian fit, respectively.}
\label{fig:ts}
\end{figure}

The LSD profiles of SZ~Psc gain a much larger SNR, compared to any individual line in the raw spectra. Figure \ref{fig:ts} shows examples of the resulting LSD profiles of SZ Psc, where we can see the wide, highly distorted profile of the rapidly rotating K star and the narrow line of the slowly rotating F star. The distortions in the K star's profiles are due to the presence of cool spots on its surface. In addition, the absorption line of the third star can also be clearly seen in the high-SNR LSD profiles, but it is much weaker than the lines of the other two components. Up to now, the properties of this third component is not well known.

Since the LSD profiles of the binary system were polluted by the third absorption features, which can lead to artefacts in the surface reconstructions, we filtered them out from the LSD profiles before the final Doppler imaging procedure. We firstly increased the errors of the affected pixels in the original LSD profiles and then performed an initial Doppler imaging with these profiles. Secondly, we fitted Gaussian profiles to the residuals of the initial solution, and subtracted them from the original LSD data set. The initial solution to the examples of the LSD profiles, the residuals and the corresponding Gaussian fits are also shown in Figure. \ref{fig:ts}. The radial velocity of the third star for each observing run is listed in the last column of Table \ref{tab:log}, and we will discuss it in Section 4.

\section{Doppler imaging}

\subsection{Surface reconstruction}

We used the imaging code DoTS \citep{cameron1992,cameron1997}, which is modified to take into account the non-synchronous rotation of the F star as described by \citet{xiang2016}, to reconstruct the surface maps of SZ~Psc. DoTS uses a two-temperature model in the description of the stellar surface, which consists of two components, a hot photosphere and a cool spot. It thus reconstructs the spot filling factors, which represent the fractional coverage of the cool spot on each individual pixel, and can not recover hot features. The iterative maximum entropy regularization method \citep{skilling1984} is implemented in DoTS to solve the ill-posed inversion problem.

Accurate stellar parameters are essential for a reliable surface reconstruction. In the Doppler imaging procedure, we adopted the orbital parameters of SZ~Psc derived by \citet{eaton2007} from spectroscopic and photometric data. We found that those parameters can well fit our observations. We also used the imaging code to fine-tune the projected equatorial rotational velocities (\vsini) of two stars and the radial velocity of the mass center of the binary system ($\gamma$) with the $\chi^2$ minimization method. The adopted parameters in Doppler imaging of SZ Psc are shown in Table \ref{tab:par}, except for $\gamma$. $\gamma$ is varying due to the wide orbital motion, and thus should be determined for each data set. The values are listed in the sixth column of Table \ref{tab:log}.

\begin{deluxetable}{lcc}
\tabletypesize{\scriptsize}
\tablecolumns{3}
\tablewidth{0pt}
\tablecaption{Adopted parameters of SZ Psc for the surface reconstruction. The letters F and K represent the F and K stars, respectively. The K star is tidally locked, so the period in the table represents both its rotational and orbital period.}
 \label{tab:par}
\tablehead{
 \colhead{Parameter}& 
 \colhead{Value}&
 \colhead{Ref.}\\
}
\startdata
 $q=M_{K}/M_{F}$ & 1.40 & a\\
 $K_{F}$ (km s$^{-1}$) & 103.98 & a\\
 $K_{K}$ (km s$^{-1}$) & 74.2  & a\\
$i$ (\degr) & 69.75 & a\\
 T$_{0}$ (HJD) & 2449284.4483 & a\\
 Period (d)  & 3.96566356 & a\\
 \vsini~$_{F}$ (km s$^{-1}$) & 3.0 & DoTS\\
 \vsini~$_{K}$ (km s$^{-1}$) & 67.7 & DoTS\\
\enddata
\tablecomments{Reference: a. \citet{eaton2007}.}
\end{deluxetable}

\begin{figure}
\centering
\includegraphics[bb = 72 18 360 596, angle=270, width=0.45\textwidth]{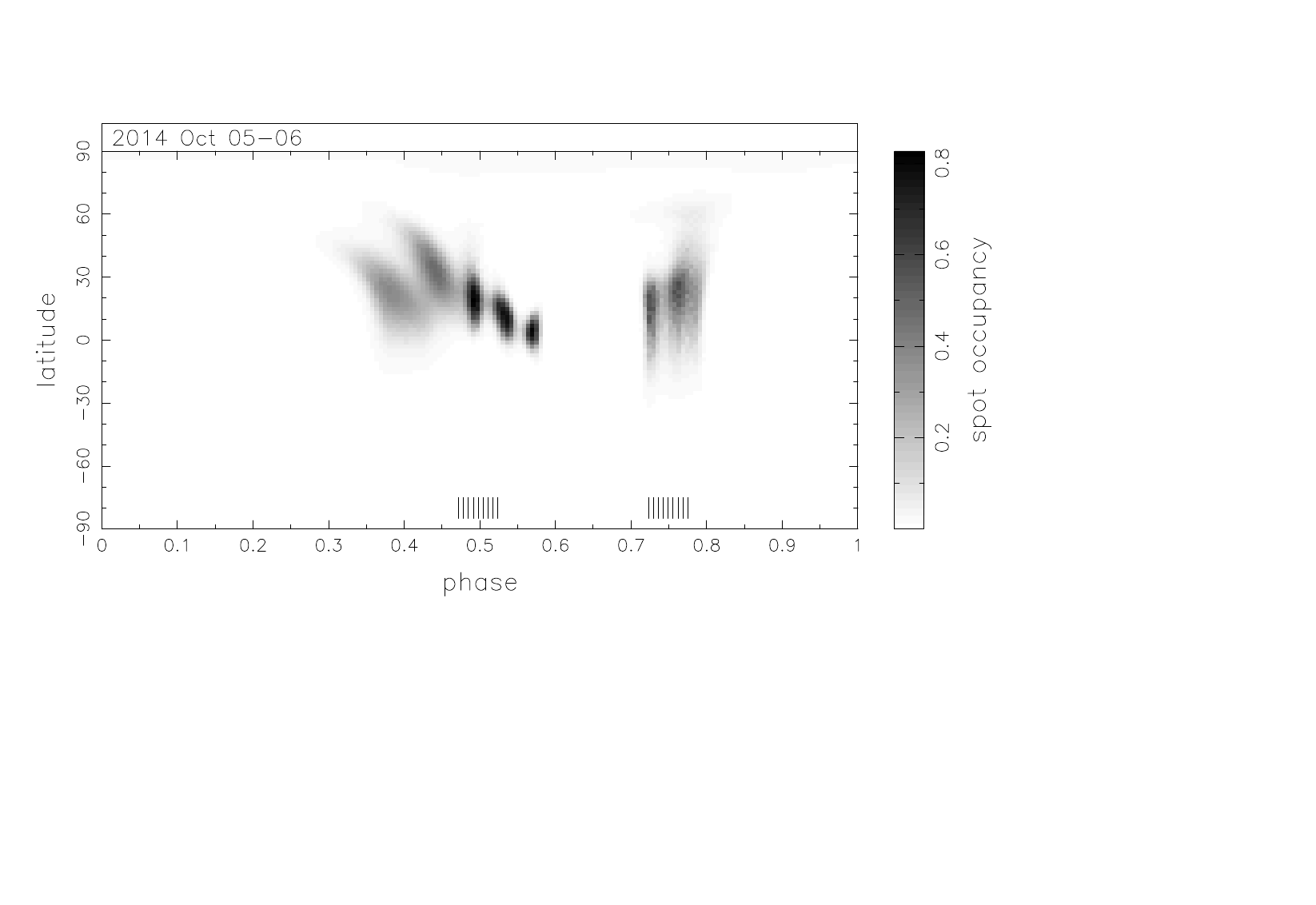}
\includegraphics[bb = 72 18 360 596, angle=270, width=0.45\textwidth]{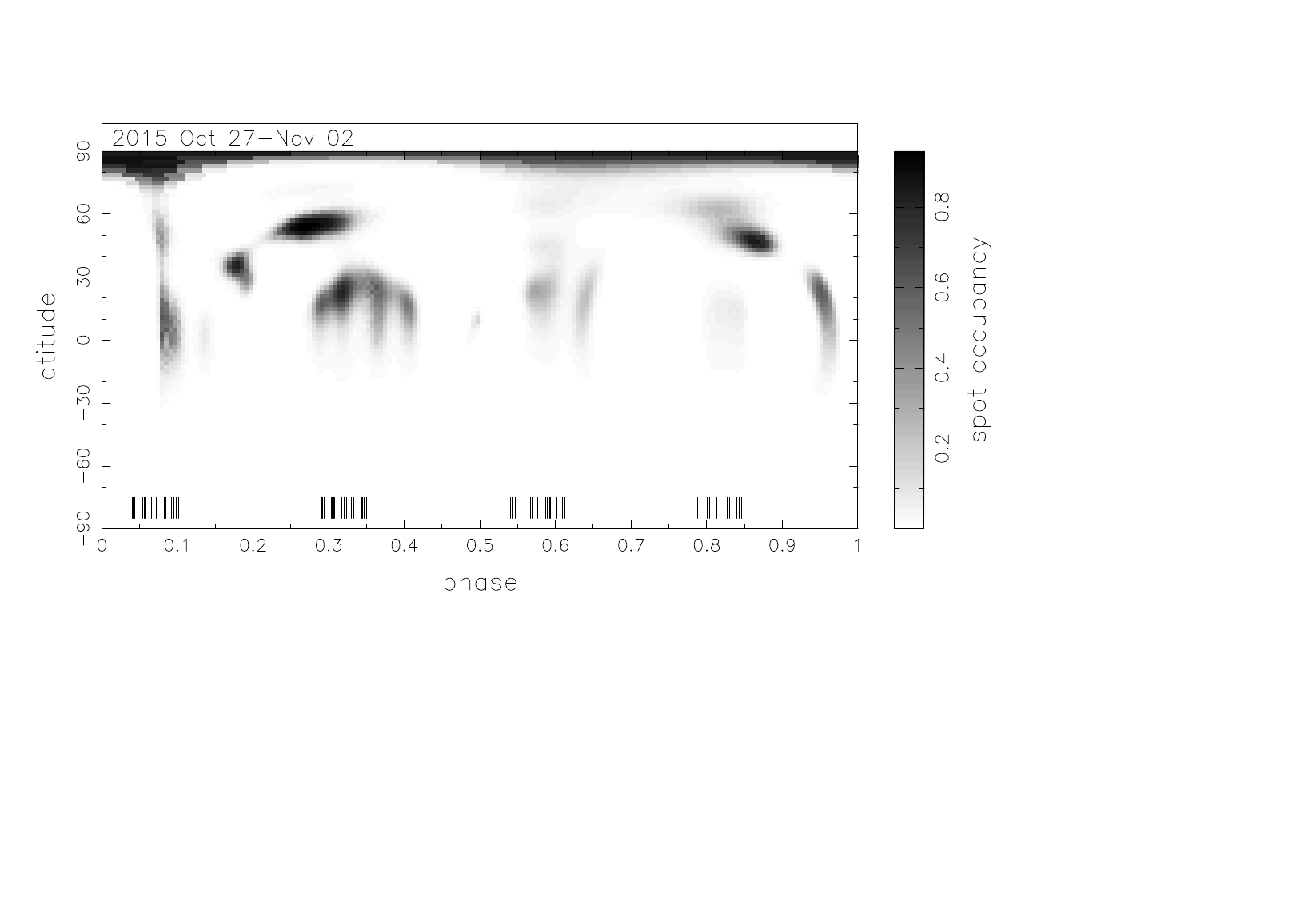}
\includegraphics[bb = 72 18 360 596, angle=270, width=0.45\textwidth]{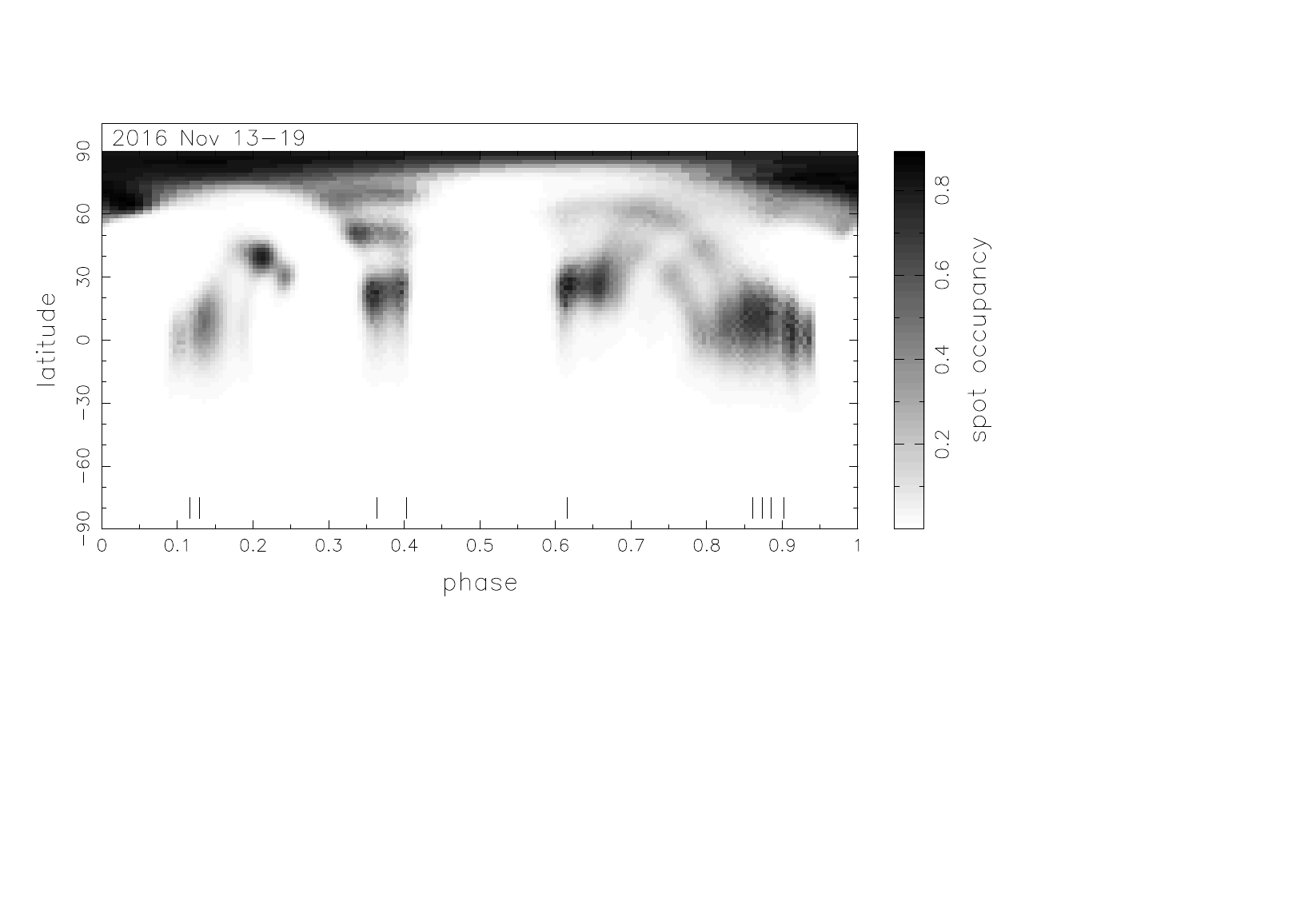}
\includegraphics[bb = 72 18 360 596, angle=270, width=0.45\textwidth]{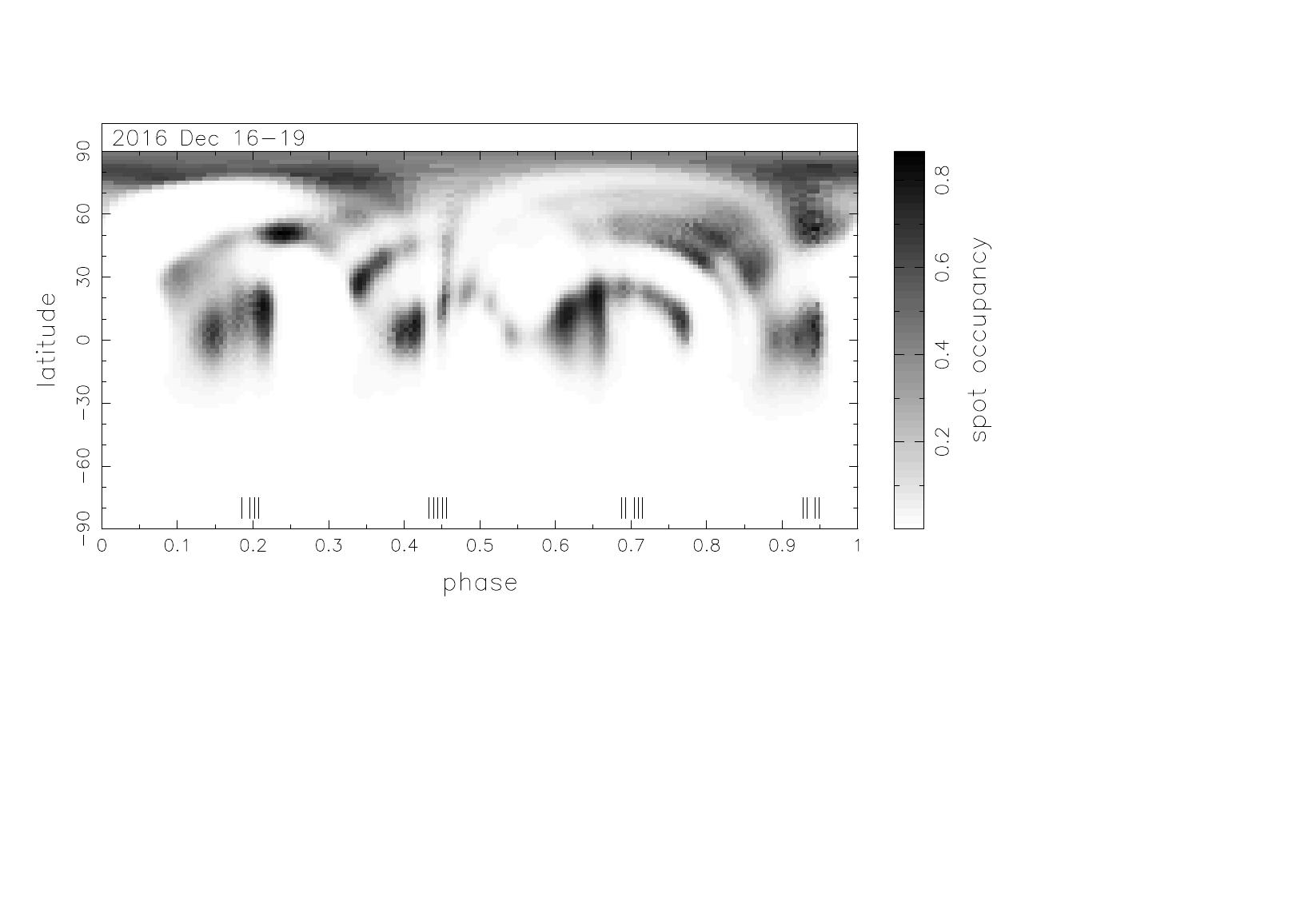}
\includegraphics[bb = 72 18 360 596, angle=270, width=0.45\textwidth]{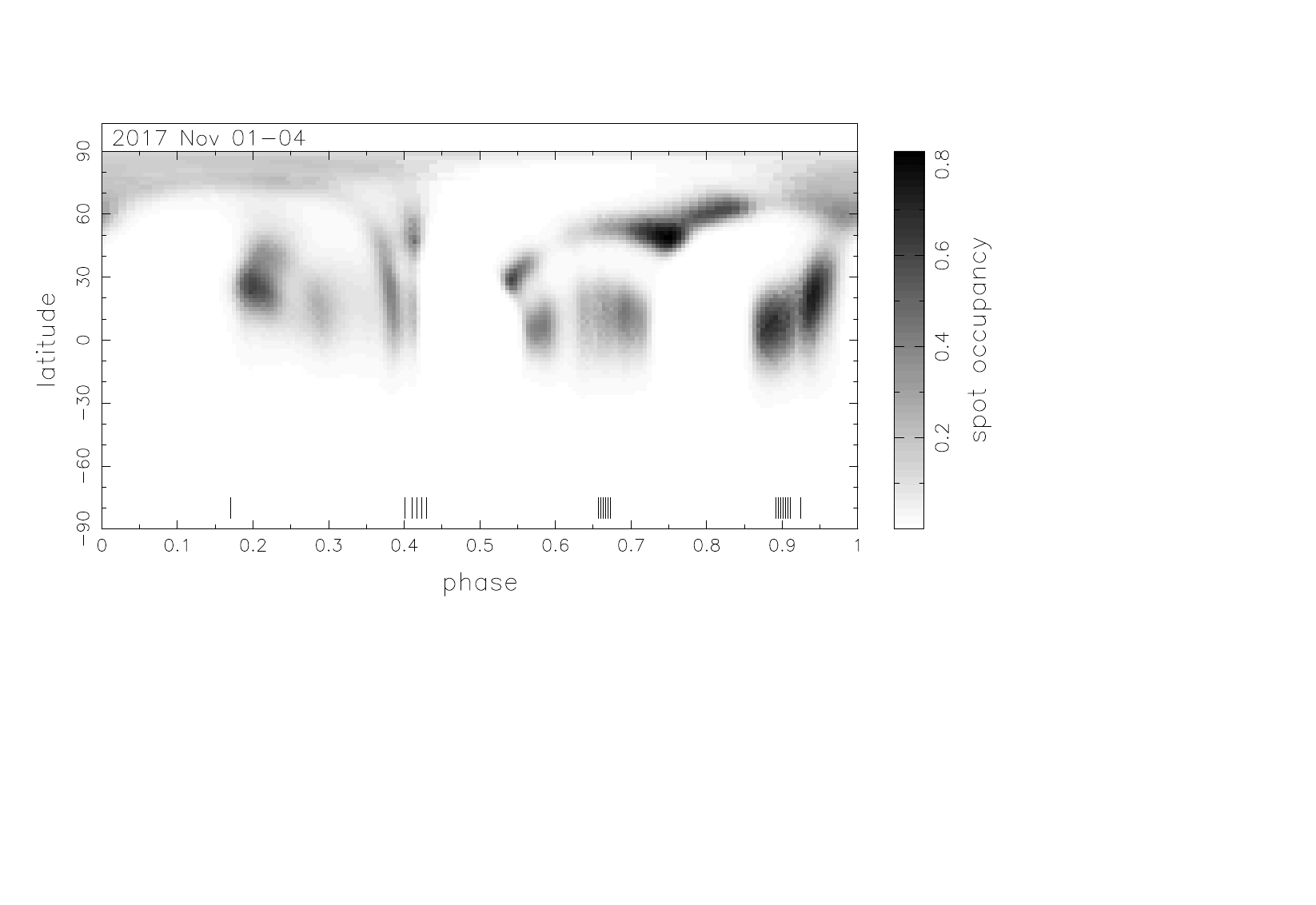}
\includegraphics[bb = 72 18 360 596, angle=270, width=0.45\textwidth]{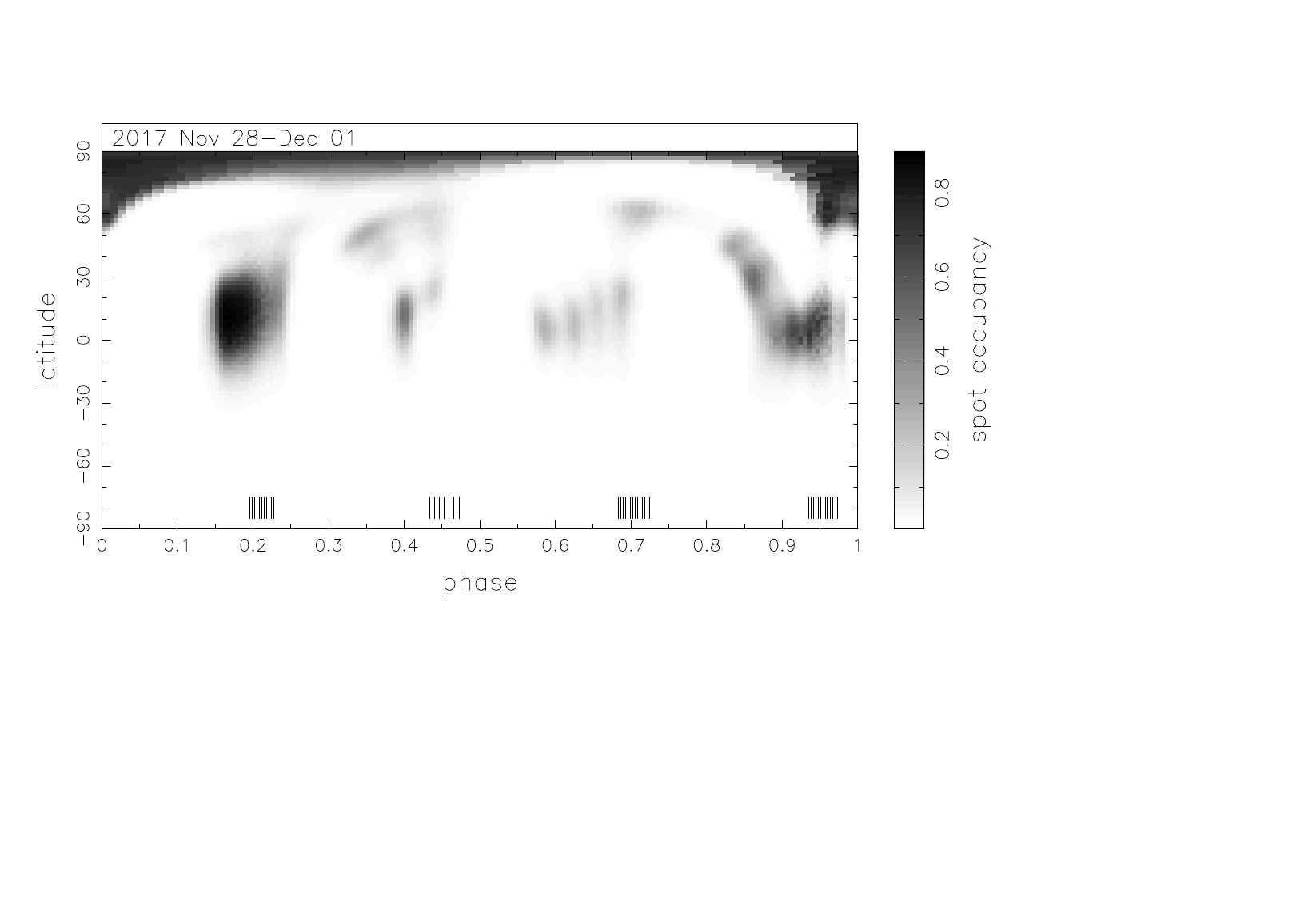}
\includegraphics[bb = 72 18 360 596, angle=270, width=0.45\textwidth]{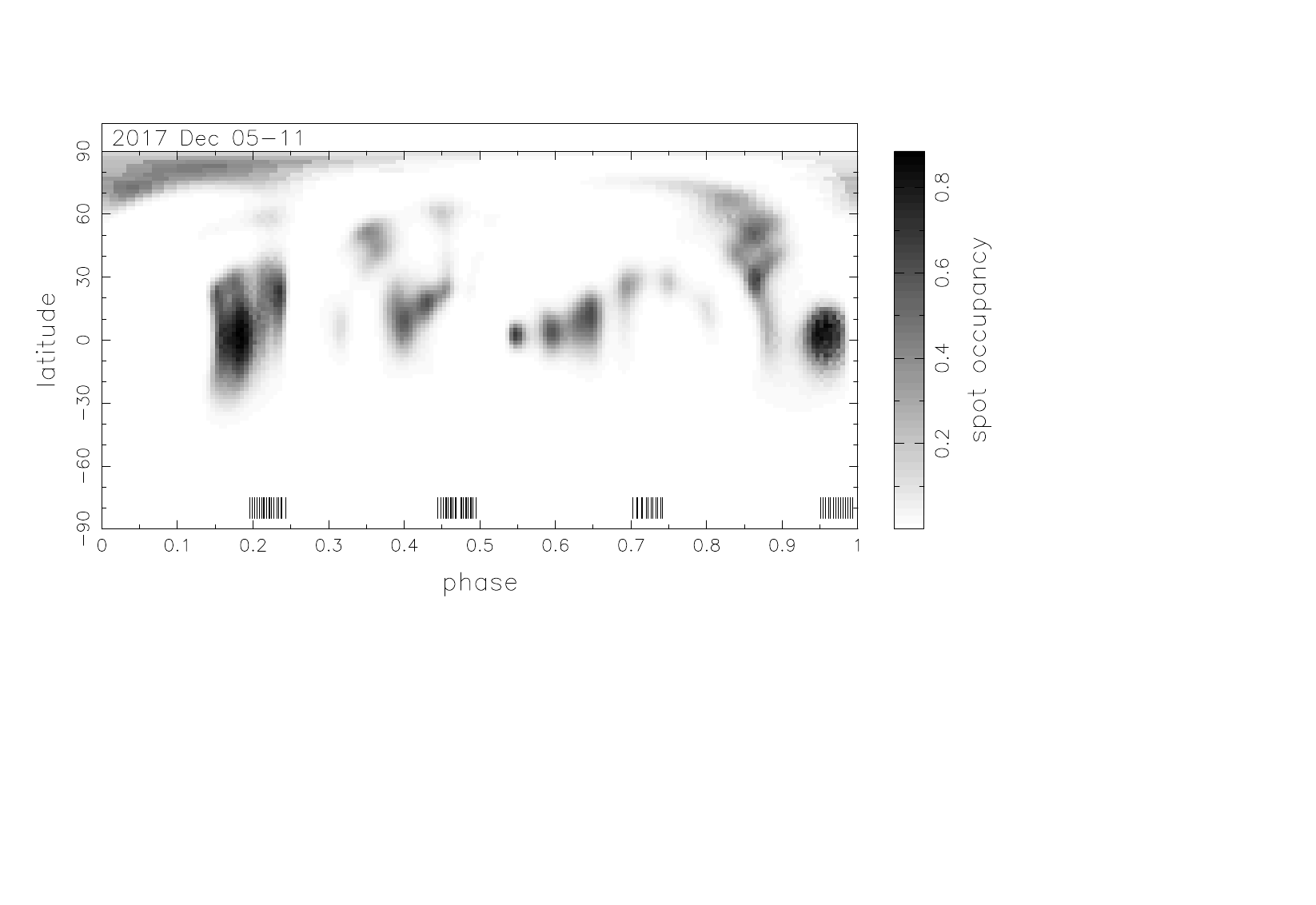}
\includegraphics[bb = 72 18 360 596, angle=270, width=0.45\textwidth]{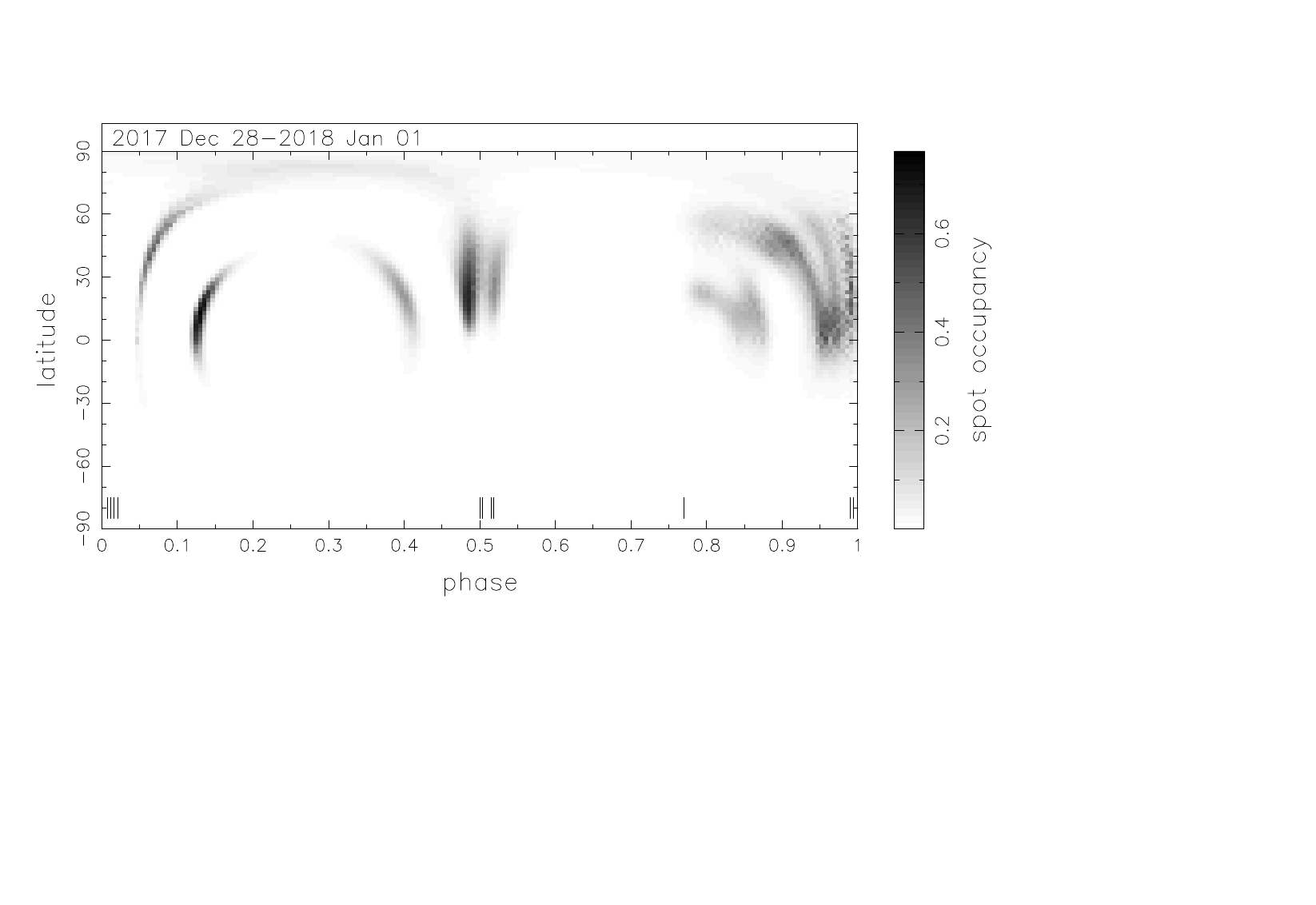}
\includegraphics[bb = 72 18 360 596, angle=270, width=0.45\textwidth]{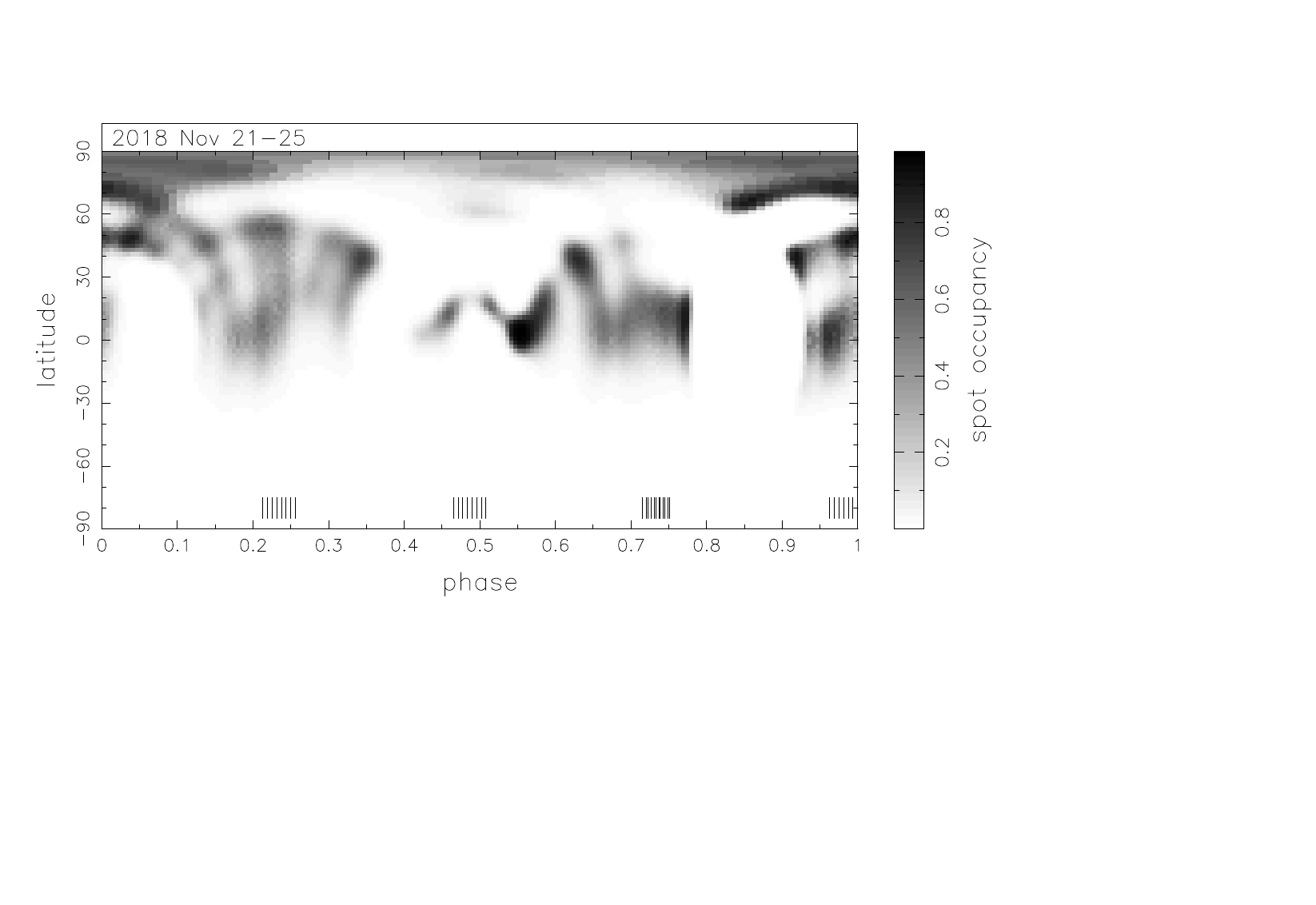}
\caption{Surface reconstructions of the K component of SZ Psc for 2014--2018. The observed phases are marked as the vertical ticks at the bottom of each images.}
\label{fig:image}
\end{figure}

We plot the modeled profiles derived by DoTS as well as the observed LSD profiles in Appendix A. The maximum entropy regularized reconstructions of the surface spot images of the K star are displayed in Figure \ref{fig:image}. In our Doppler images, phase 0.5 on the K star is facing the F star. Given the spectral resolution of 48000 and the \vsini\ of the K star of 67.7 km/s, we can estimate that the longitude resolution of our Doppler imaging maps is about 8\degr\ at the equator. The latitude resolution of Doppler imaging is poor around equator, which causes the low-latitude spots to appear vertically elongated \citep{cameron1994}.

The image of 2015 Oct shows a number of spots at various latitudes, including a non-axisymmetric polar spot with appendages extending to the equator at phase 0.1, two intermediate-latitude (30\degr--60\degr) spots at phase 0.3 and 0.9, and some low-latitude (0\degr--30\degr) spot groups. The appendages are probably artefacts smeared from the real spots as discussed in Section 3.2. The both images of 2016 are similar to each other. They both show polar spot with appendages at phase 0.4 and several low- to intermediate latitude spots. But the positions and strength of these spots changed slightly in one month. The image of 2017 Nov 01--04 shows spot structures around phases 0.3, 0.7 and 0.9. The spot maps of 2017 Nov 28--Dec 01 and 2017 Dec 05--11 are also similar to each other, since they were taken at very close epochs. They show polar spots at phase 0 and low latitude spots at phases 0.2, 0.4, 0.6 and 0.9. The surface map of 2018 shows large low-latitude spot groups around phases 0, 0.2 and 0.7. The 2014 image shows only two low-latiude spot groups at phases 0.5 and 0.75, and the image of 2017 Dec 28--2018 Jan 01 shows low-latitude features at phases 0.1, 0.5 and 0.9. The lack of spots in these two images should be due to the incomplete phase coverage.

\subsection{Phase coverage test}

Due to the almost integral rotational period of SZ Psc, it is very difficult to effectively obtain an ideal rotational phase coverage at a single observing site. The phase gaps of the observations can not be decreased to below 0.2 due to the visibility of SZ Psc and the daylight. We thus tested the effect of the phase coverage on our Doppler images by using the simulated data.

\begin{figure}
\centering
\includegraphics[bb = 72 18 360 596, angle=270, width=0.45\textwidth]{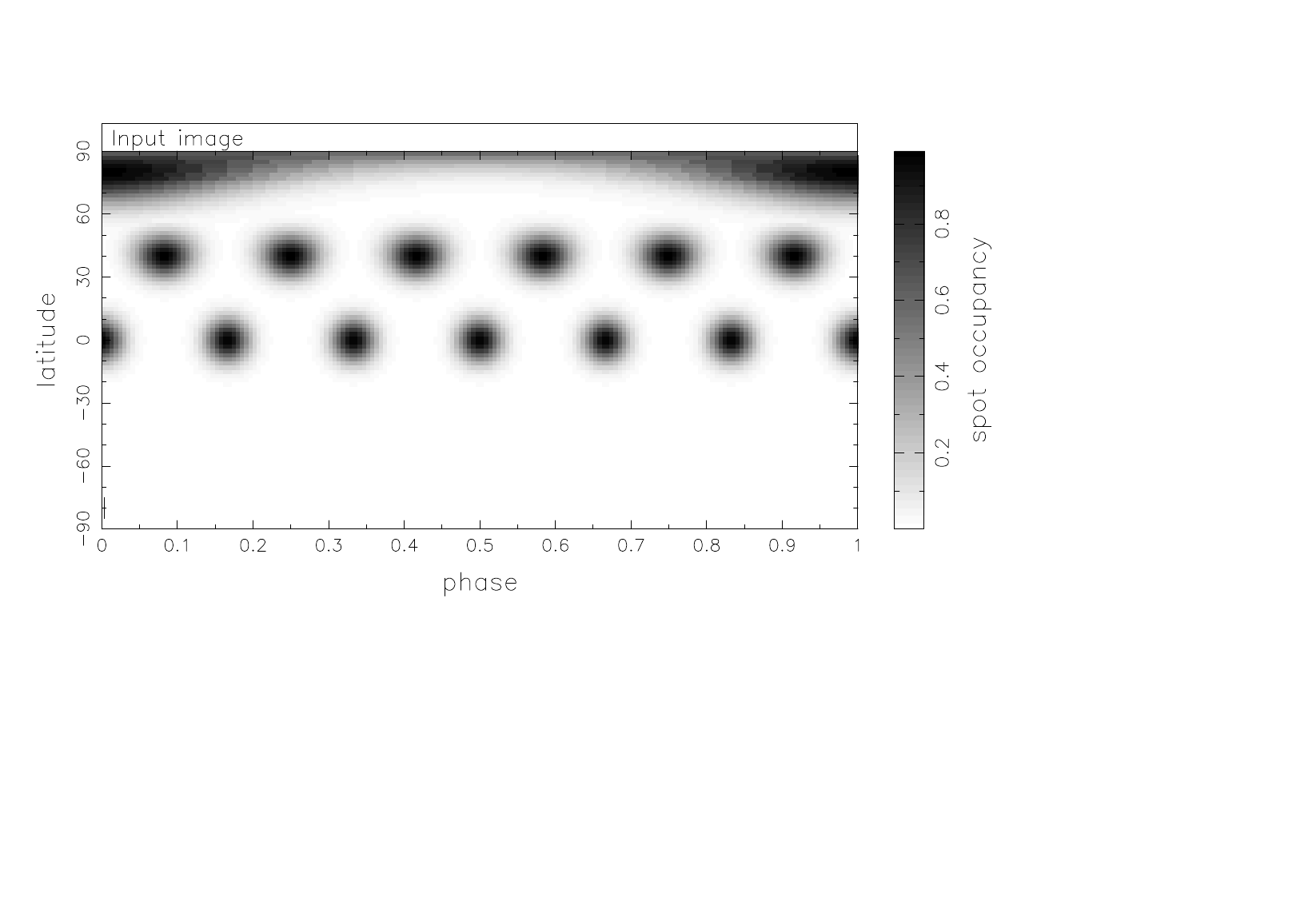}
\includegraphics[bb = 72 18 360 596, angle=270, width=0.45\textwidth]{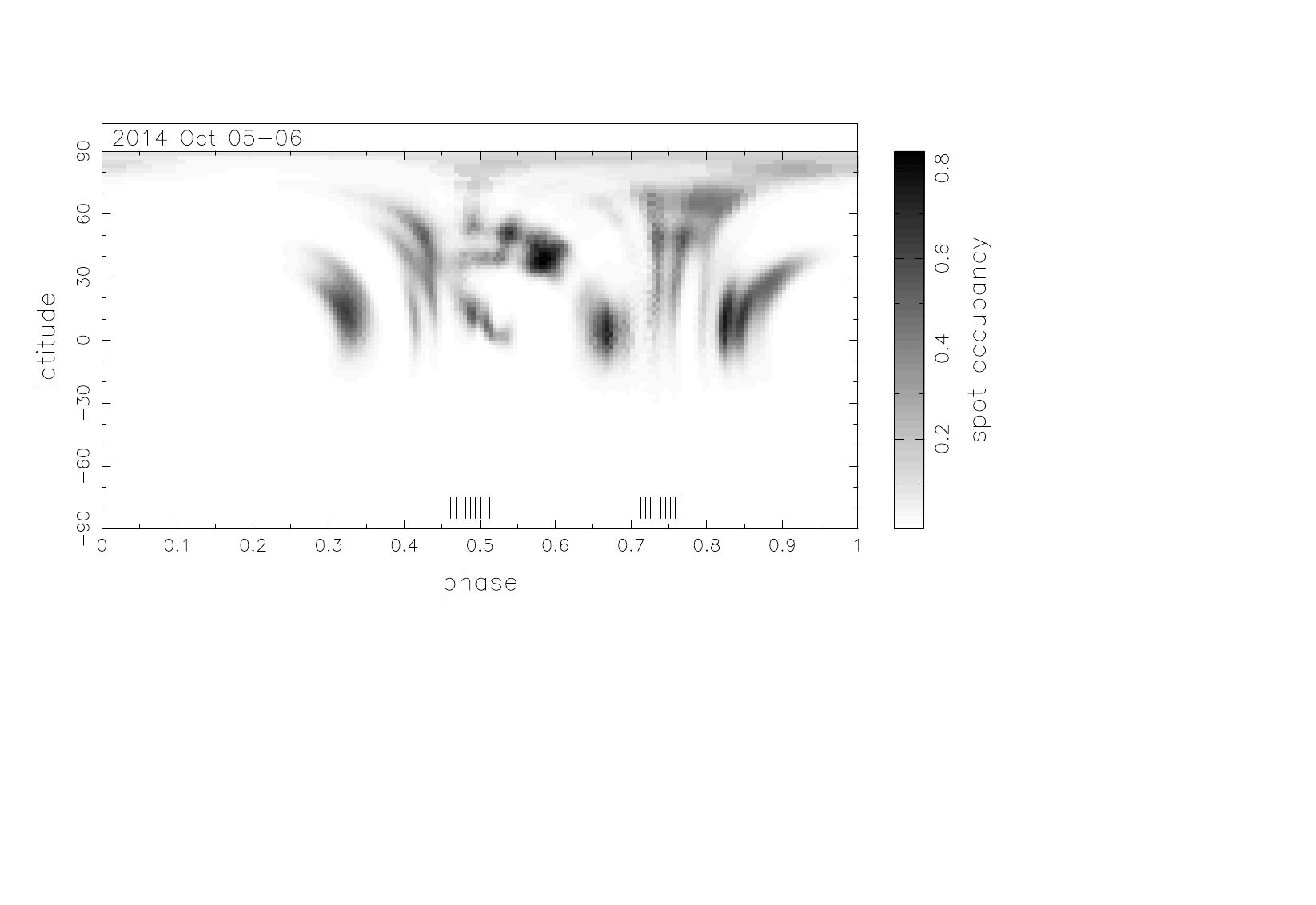}
\includegraphics[bb = 72 18 360 596, angle=270, width=0.45\textwidth]{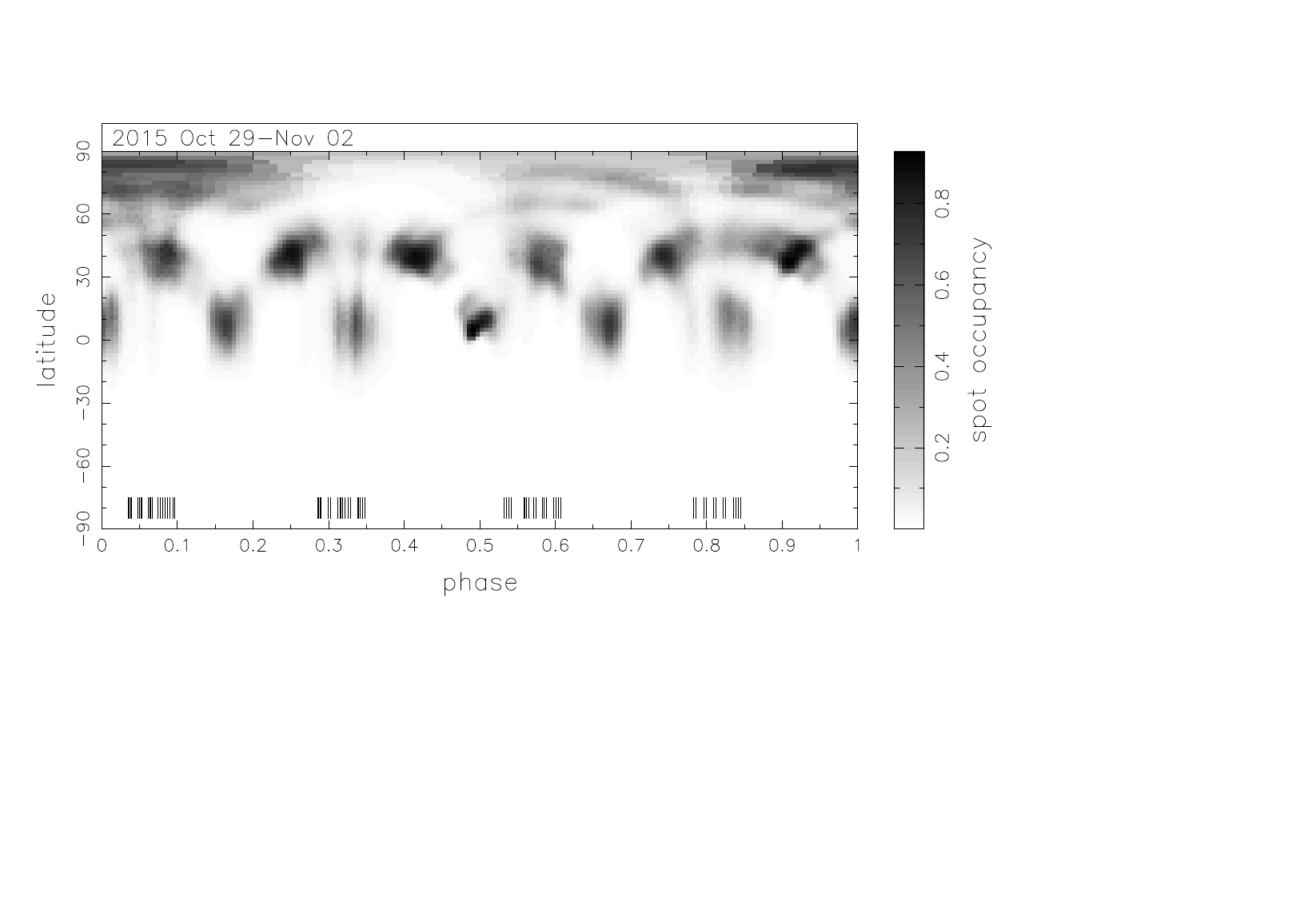}
\includegraphics[bb = 72 18 360 596, angle=270, width=0.45\textwidth]{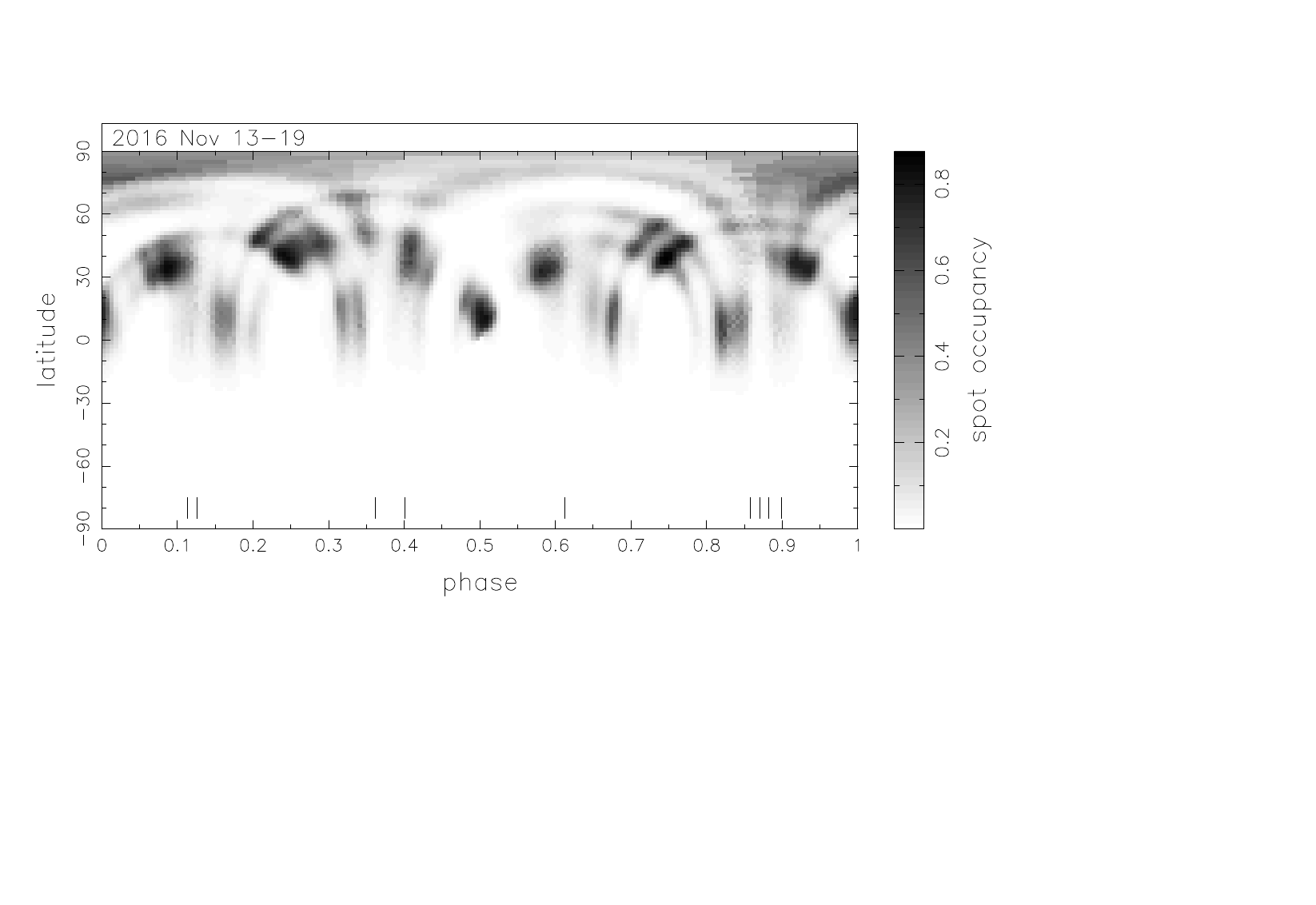}
\includegraphics[bb = 72 18 360 596, angle=270, width=0.45\textwidth]{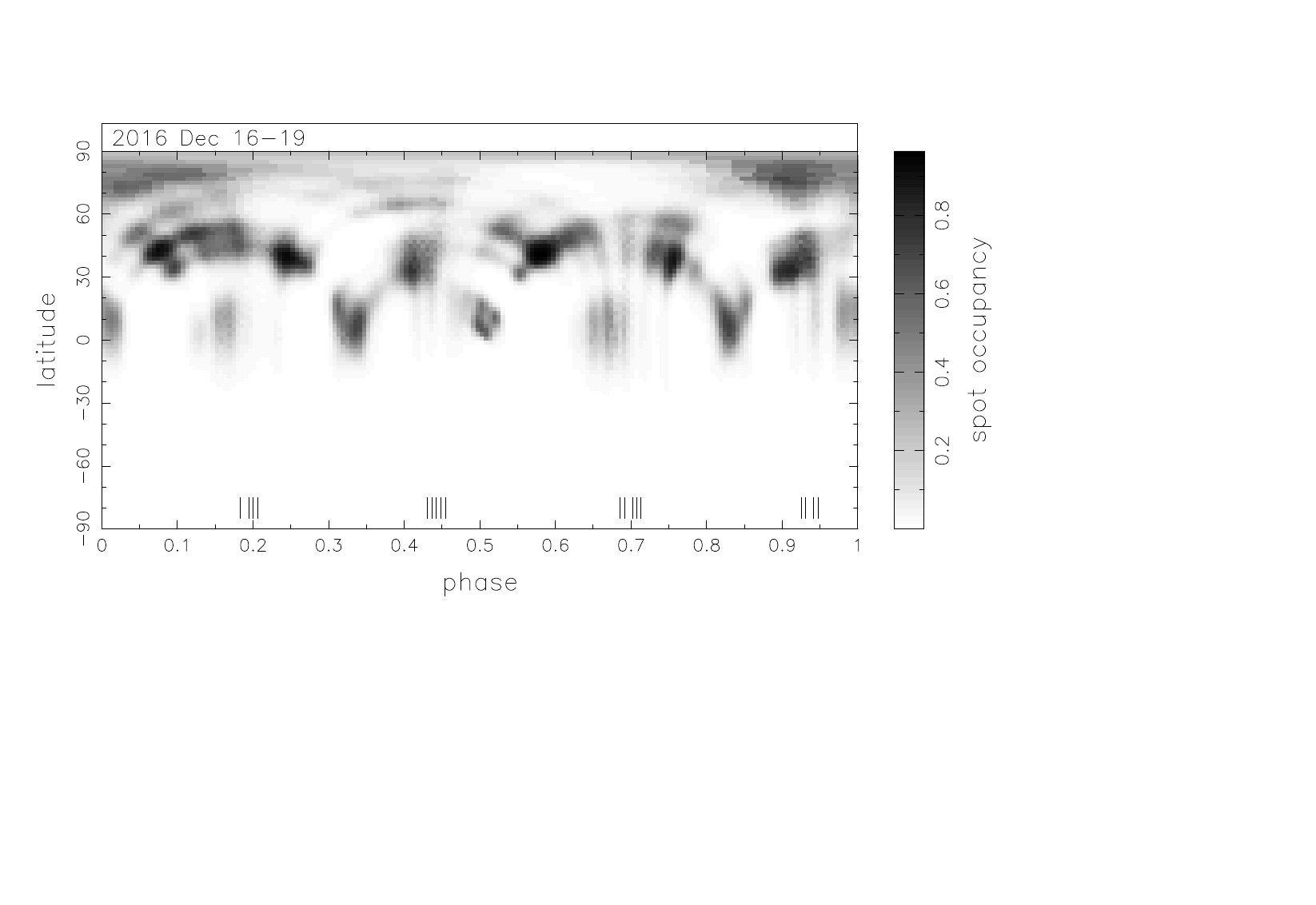}
\includegraphics[bb = 72 18 360 596, angle=270, width=0.45\textwidth]{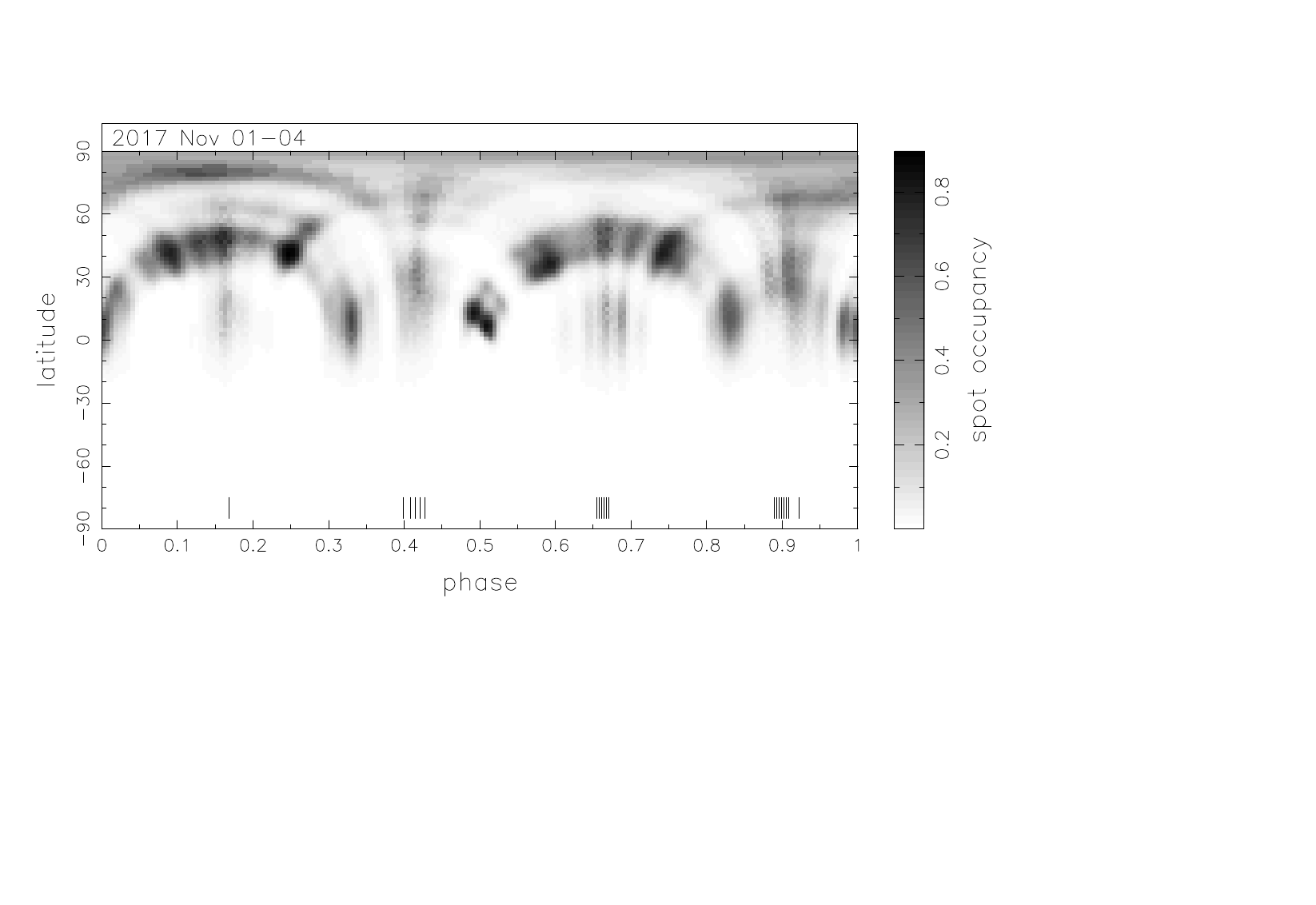}
\includegraphics[bb = 72 18 360 596, angle=270, width=0.45\textwidth]{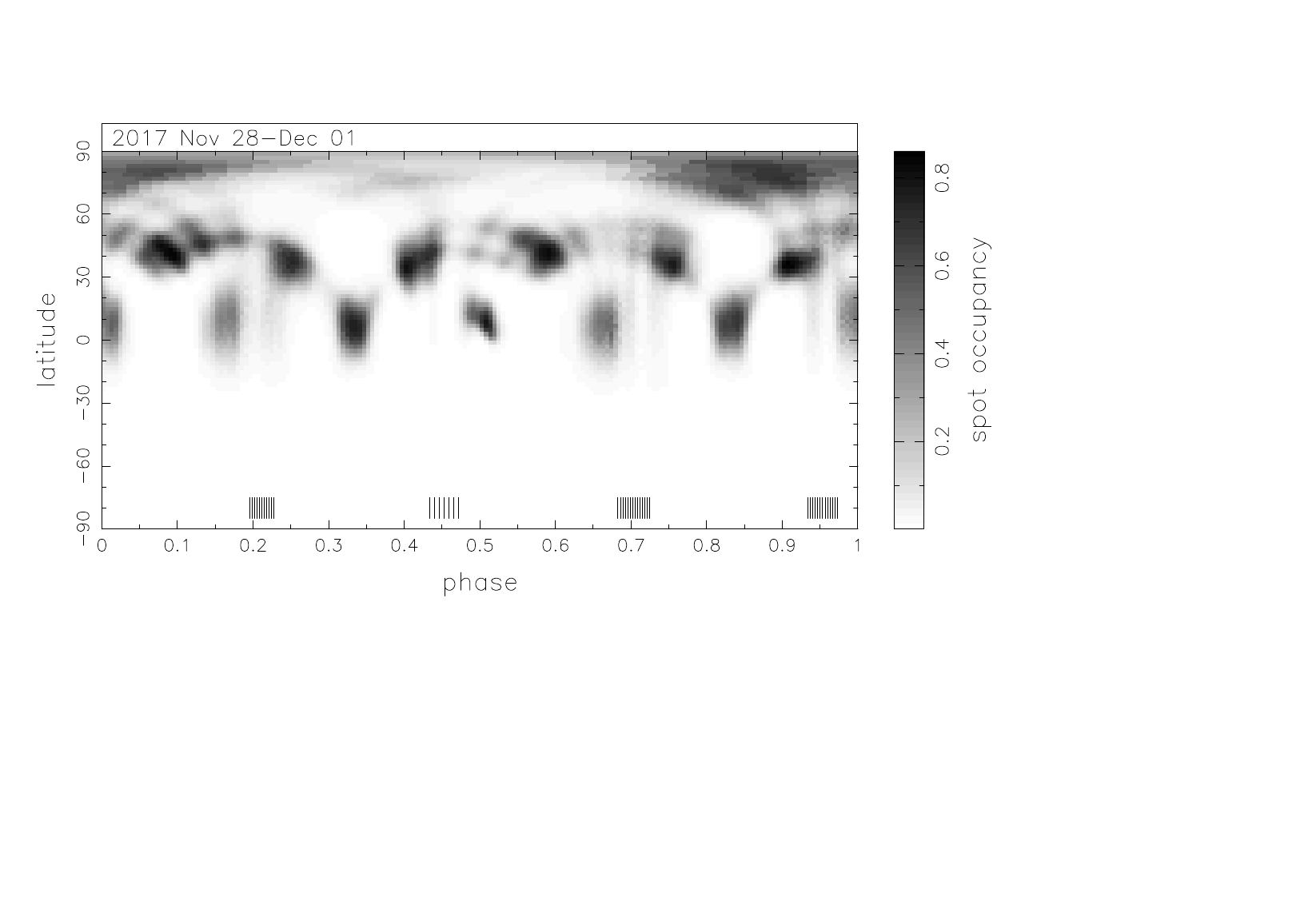}
\includegraphics[bb = 72 18 360 596, angle=270, width=0.45\textwidth]{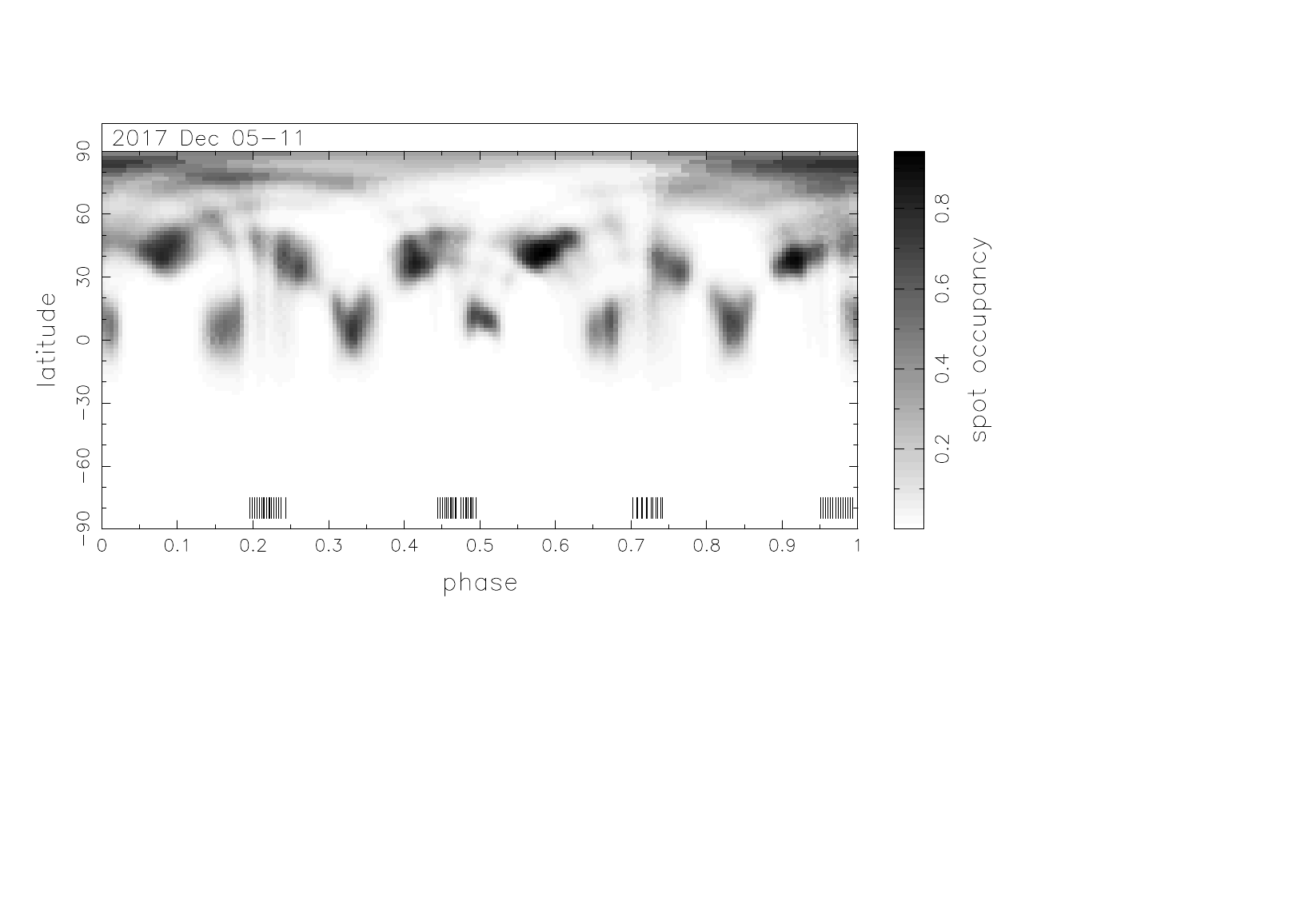}
\includegraphics[bb = 72 18 360 596, angle=270, width=0.45\textwidth]{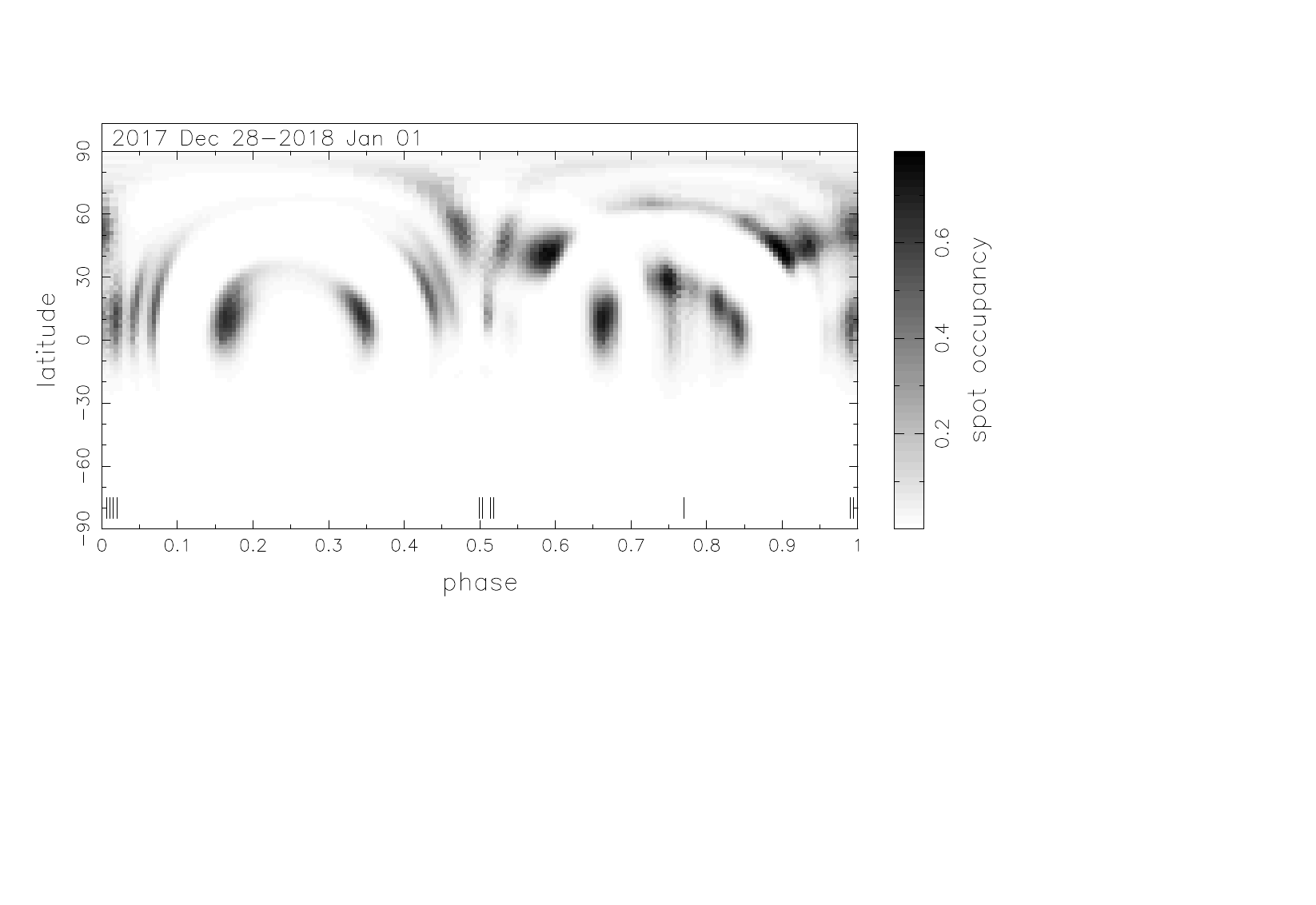}
\includegraphics[bb = 72 18 360 596, angle=270, width=0.45\textwidth]{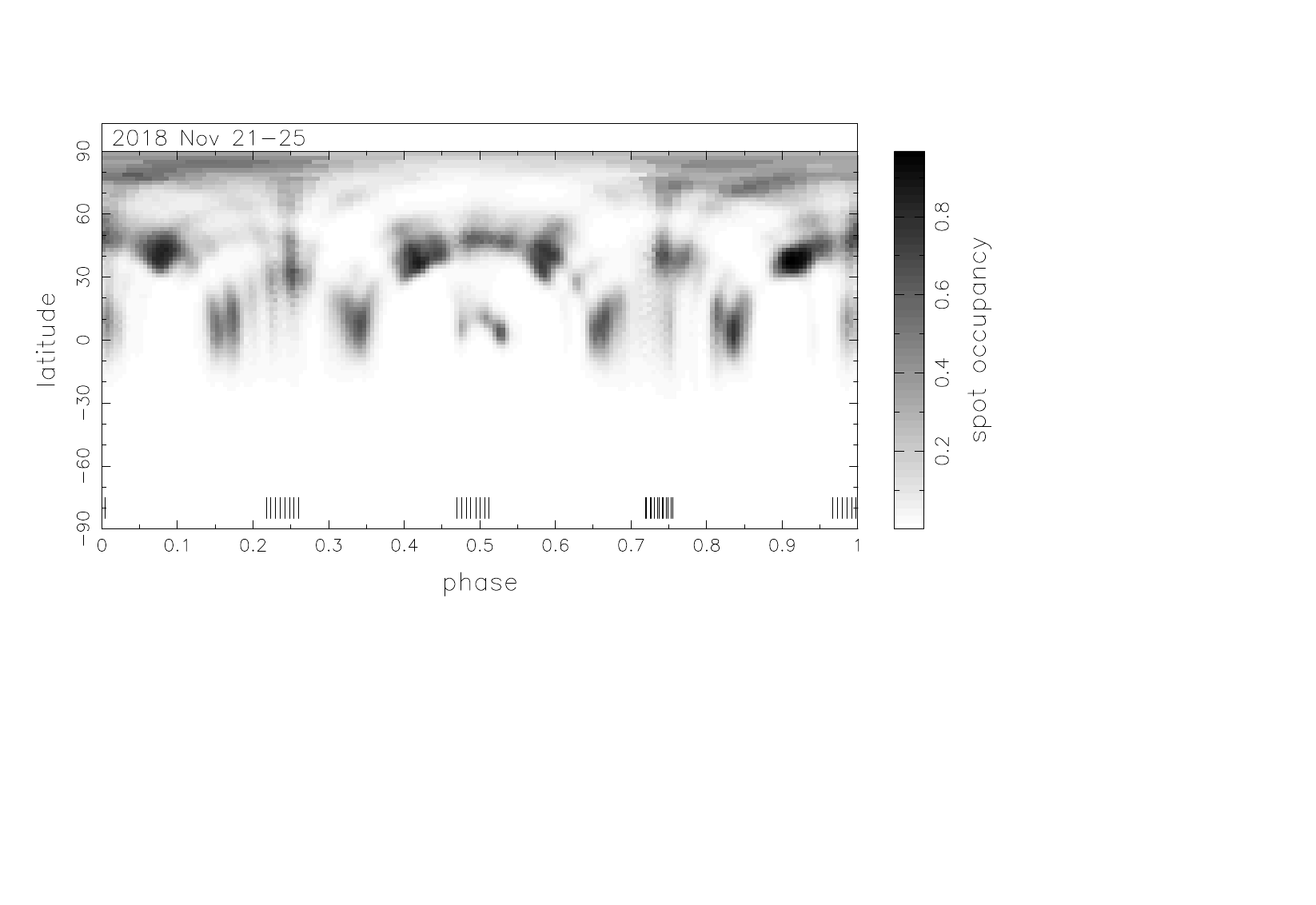}
\caption{Tests on the effect of the phase coverage with the simulated images. The first panel is the input image used to produce profiles. Others are reconstructed images from the simulated data sets with different phase sampling.}
\label{fig:test}
\end{figure}

We made a stellar image that has a non-axisymmetric polar spot at phase 0 and several evenly distributed intermediate- and low-latitude spots as the input image and used the imaging code DoTS to produce the corresponding LSD profiles with phase coverage of our different data sets. The input and reconstructed test images are shown in Figure \ref{fig:test}.

The results show that the reconstruction is reasonably reliable if the data set has observations of 4 consecutive days or longer. The large phase gaps in the simulated data of 2014 Oct and 2017 Dec produce spurious features and lead to lack of spot patterns. The spurious arc structures, smeared from real spots, are produced due to the phase gaps of about 0.2 in all of the images, which should be considered carefully.

\subsection{Differential rotation}

Stellar magnetic field of late-type stars is produced by the interplay between rotation and convection motions of the plasma in the stellar convection zones. Differential rotation is one of the key ingredients of the dynamo process, which generates the stellar magnetic field. The cross-correlation of the Doppler images in short time can be used to measure the stellar surface flow \citep{donati1997b,weber2001}.

The 2017 Nov and Dec datasets were collected in near epochs, which provides an opportunity for us to firstly determine the surface differential rotation of SZ Psc. Here, we did not directly use the image derived from the data set of 2017 Dec 05--Dec 11, spanning 6 days, but reconstructed a new image using the data of 2017 Dec 08--Dec 11 for the measurement of the differential rotation, so that both images were from 4-consecutive-day observations, 2017 Nov 28--Dec 01 and 2017 Dec 08--Dec 11, which are about 10 days apart. The observations in 4 consecutive days can provide sufficient phase coverage, as discussed in Section 3.2.

We calculated the cross-correlation function (CCF) of each latitude between the Doppler images of these two epochs in 2017 Nov and Dec, and then fitted a solar-like differential rotation law, which is described below, to the peaks of the CCF of latitude bands.
\begin{equation}
\Omega (\theta) = \Omega_{eq} - \Delta \Omega \sin^{2} \theta,
\end{equation}
where $\Omega (\theta)$ and $\Omega_{eq}$ are the rotational rates at latitude $\theta$ and the equator, respectively, and $\Delta \Omega$ is the difference of angular velocities between the equator and the pole. A positive $\Delta \Omega$ means a solar-like differential rotation. Due to the relatively weak longitude resolution near the rotational pole and low latitude resolution near the stellar equator, we only used the values between latitudes 20 and 75 degrees.

\begin{figure}
\centering
\includegraphics[width=0.66\textwidth]{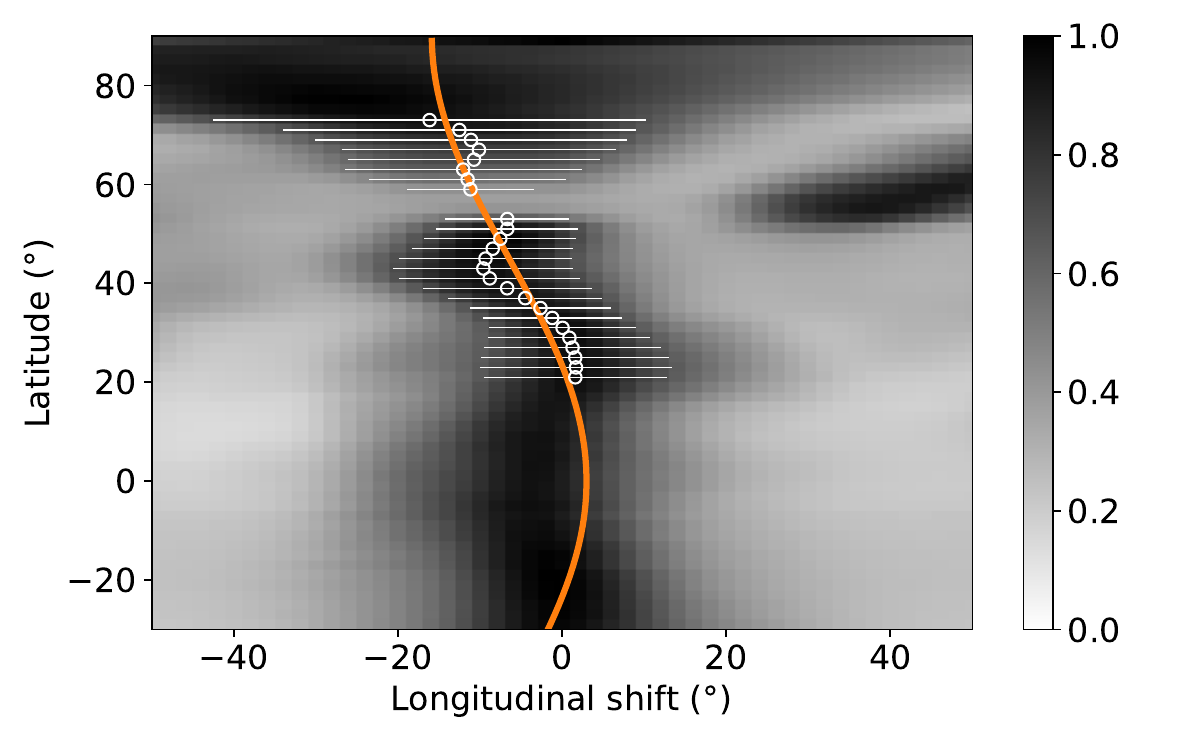}
\caption{Cross-correlation functions of each latitude band for the K star of SZ~Psc. The peak (white open circle) of the cross-correlation function for each latitude is derived by fitting a Gaussian profile and the full width at half maximum (FWHM) of the profile is plotted as the white horizontal line. The orange curve is the fit of the solar-like differential rotation law.}
\label{fig:dr}
\end{figure}

The CCF distribution map and the fitted surface shear curve are both shown in Figure \ref{fig:dr}. The estimated differential rotation parameters of the K component are $\Omega_{eq} = 1.591 \pm 0.002$ rad d$^{-1}$ and $\Delta \Omega = 0.035 \pm 0.003$ rad d$^{-1}$. The K star shows a solar like differential rotation, the equator area rotates faster than the rotational pole area.

\section{Orbit of the third star}

The existence of a third component of SZ~Psc was previously reported by \citet{eaton2007} and \citet{glazunova2008} from spectra line profiles. In our last work on Doppler imaging study of SZ~Psc \citep{xiang2016}, we used the spectra obtained in 2004 and 2006 to determine the radial velocities of the third star of this triple system and found a period of 1280 days, assuming a simple circular orbit. But the observations were only made in two years and the number of the observed LSD profiles were too limited to cover the entire wide orbital phase.

\begin{figure}
\centering
\includegraphics[width=0.66\textwidth]{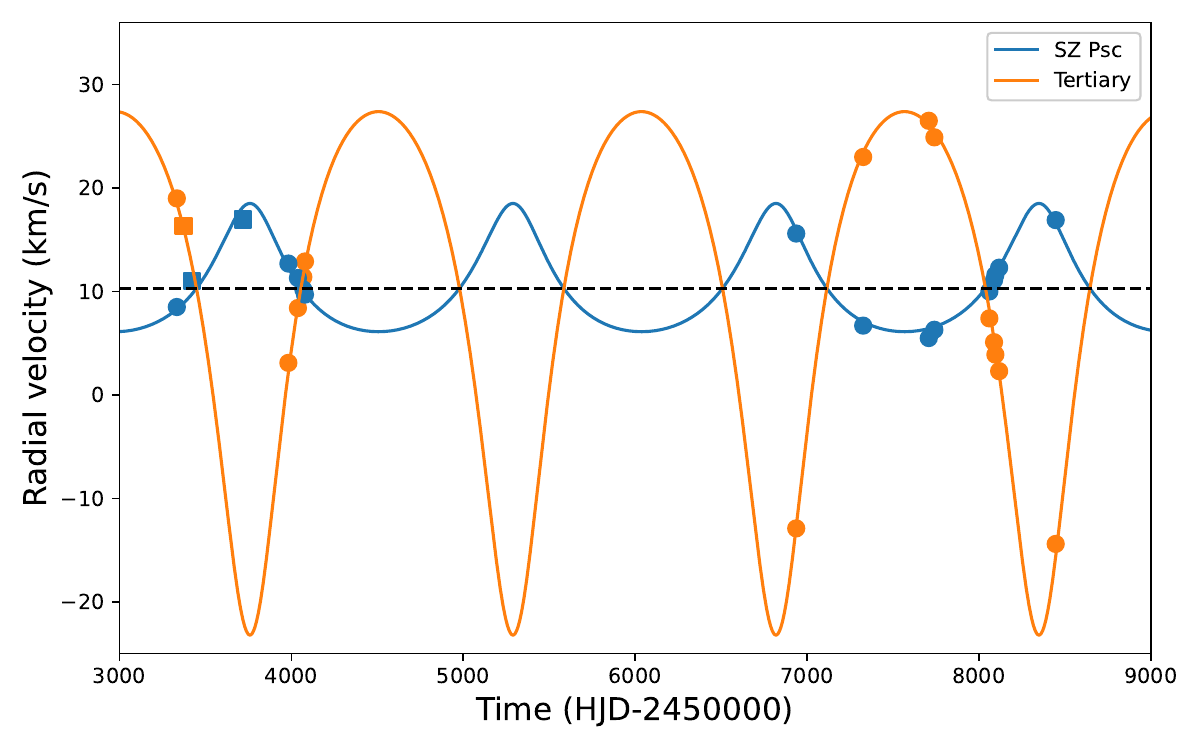}
\includegraphics[width=0.66\textwidth]{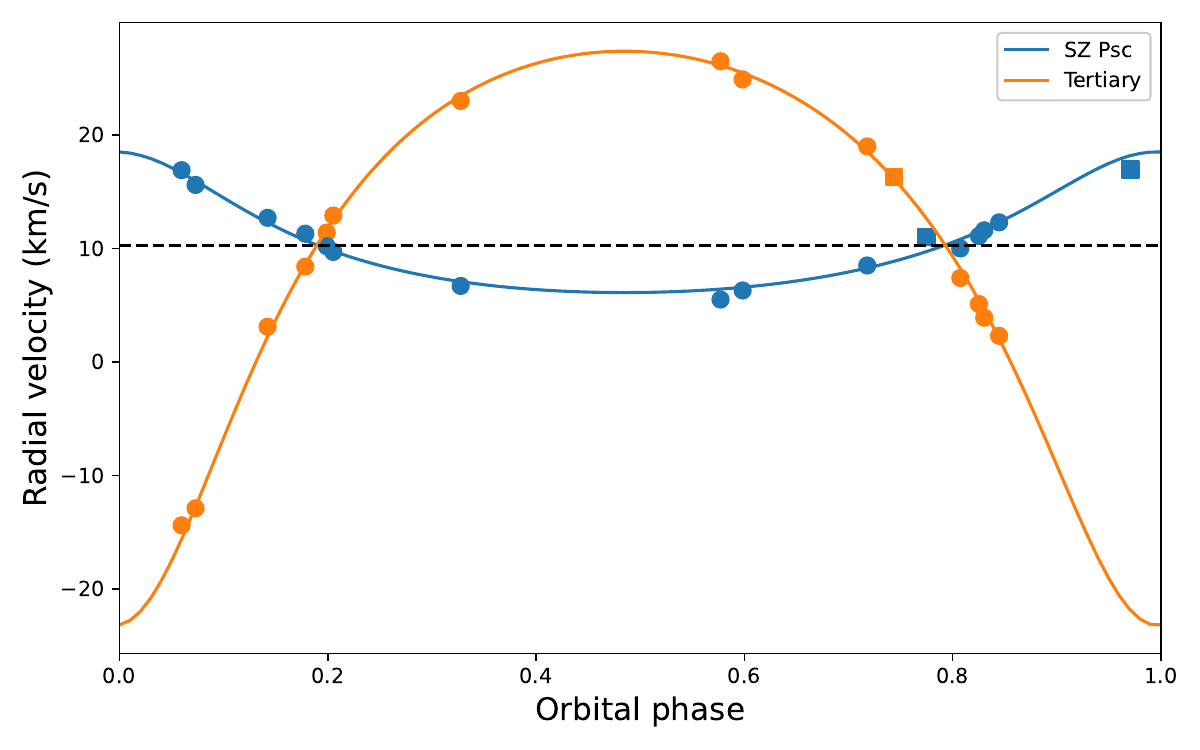}
\caption{Raw (upper) and phase folded (bottom) radial velocity curves of the binary system (blue) and the third star (orange). The dots represent our measurements. The blue squares are values determined by \citet{eaton2007} and the orange square represents the value derived by \citet{glazunova2008}.}
\label{fig:rv}
\end{figure}

In this work, we also measured the radial velocity of the third component from the new LSD profiles for each data set, and corrected it to the heliocentric frame. Combined with the previous data, we fitted the radial velocities of the binary system and the third component simultaneously, as shown in Figure \ref{fig:rv}. As a result, we derived that the period of the outer orbit of the triple system is $1530 \pm 3$ days. \citet{eaton2007} derived the total mass of the binary system to be 3.07 M$\odot$. According to the amplitudes of two radial velocity curves, we estimated the mass of the third star to be $0.75 \pm 0.06$ M$\odot$. The parameters of the outer orbit are listed in Table \ref{tab:orb}.

\begin{deluxetable}{lcc}
\tabletypesize{\scriptsize}
\tablecolumns{2}
\tablewidth{0pt}
\tablecaption{Orbital parameters of the triple system. P is the period of the outer orbit of SZ~Psc, $e$ is the eccentricity, $\omega$ is the argument of periapsis, T$_{0}$ is the time of periapsis passage and $\gamma$ is the radial velocity of the mass center of the triple system.}
 \label{tab:orb}
\tablehead{
 \colhead{Parameter}& 
 \colhead{Value}&
}
\startdata
 $P (d)$ & 1530 $\pm$ 3\\
 $e$ & 0.325 $\pm$ 0.013\\
 $\omega$ (\degr) & 3.0 $\pm$ 1.7 \\
 T$_{0}$ (HJD) & 2452235 $\pm$ 13\\
 $K_{\textrm{Binary}}$ (km/s) & 6.2 $\pm$ 0.3\\
 $K_{\textrm{Tertiary}}$ (km/s)& 25.3 $\pm$ 0.8\\
 $\gamma$ (km/s)& 10.3 $\pm$ 0.2\\
\enddata
\end{deluxetable}

\section{Discussion and conclusions}

We have presented a series of 9 new Doppler images of the magnetically active K subgiant component of the eclipsing RS CVn-type binary SZ Psc, based on the high-resolution spectra collected from 2014 to 2018. The K subgiant always shows a wide distribution of starspots along its latitude and longitude. The combined effect of the Coriolis force and meridional flow in the convection zone can lead to the formation of starspots across a wide range of latitudes on rapidly rotating stars \citep{schrijver2001,holzwarth2006}. The complexity of starspot patterns is very similar to that revealed by our previous Doppler images \citep{xiang2016}. The starspot patterns consist of a number of spot groups. The spot configurations on the K star revealed by Doppler imaging are consistent to the result of \citet{eaton2007}, who found that more than 15 spots are required to explain the line distortions.

A prominent feature on the K subgiant is a polar starspot that is asymmetric with respect to the rotational pole. It has been uncovered by all of the images with sufficient phase coverage. In our previous Doppler imaging study of SZ Psc \citep{xiang2016}, the surface image of 2006 Nov-Dec, which had a good phase sampling, showed a similar non-axisymmetric polar/high-latitude feature. It seems that this polar spot is a stable structure on the K subgiant. Large polar starspot is a common feature on the surface of fast rotators, discovered by both of Doppler imaging and interferometry \citep{roettenbacher2016}.

From the comparison between two surface images of 2016 Nov 13--19 and Dec 16--19, we can see starspot evolution in a time-scale of one month. The polar structure was relatively stable, but the low- to intermediate-latitude spot groups evolved over time. The intermediate latitude appendages of polar spot at phase 0.4 became much extended in one month. New weak features emerged around phase 0.5 but they may be artefacts smeared from the nearby spots. \citet{cao2020} revealed that the chromosphere activity of SZ Psc changed within one month in 2016 Nov and Dec, using the same data sets. From long-term light curves, \citet{giles2017} and \citet{namekata2019} showed that the lifetime of starspots on solar-like stars spans 10--350 days, and is related to the starspot areas, effective temperature and the rotational periods. From long-term photometry, \citet{henry1995} determined the lifetimes of spots on RS CVn binaries, ranging from months to years. \citet{namekata2020} traced the evolution of individual spots on the Sun-like star Kepler-177 by analyzing spot-crossing events and found that it is different from the evolution derived through the rotational modulation, which reveals that the evolutions of the spatially resolved spots differ from those of the starspot groups or active longitudes. \citet{lanza2001} revealed the change of light curve of SZ Psc is dominated by the starspot evolution.

\citet{cao2020} revealed active longitudes at phases 0.25 and 0.75 on the K component through the analysis of the chromospheric activity indicators. The longitudinal distribution of starspots on the K star is always non-uniform, but we can not find the consistent active longitudes. The tidal force affects the distribution of starspots on binary stars to form preferred longitudes. For instance, the very rapidly rotating close binary ER Vul (P=0.7 days) shows starspots concentrated on the hemispheres facing each other on both components \citep{xiang2015}. The starspots on the K star of SZ Psc show a complicated distribution instead of one or two active regions.

In order to validate our Doppler images with different phase coverage, we have also performed the tests on the simulated image and profiles. The results show that the data set spanning at least 4 consecutive days has an enough phase coverage and can produce a reliable reconstruction.

We have measured the surface differential rotation rate of the K subgiant from the cross-correlation function of two Doppler images about 10 days apart. The parameters are $\Omega_{eq} = 1.591 \pm 0.002$ rad d$^{-1}$ and $\Delta \Omega = 0.035 \pm 0.003$ rad d$^{-1}$, which can be translated into an equator-pole lap period of 180 days on the K star. It is about 1.5 times longer than that on the Sun. The relative differential rotation coefficient is $\alpha = \Delta \Omega / \Omega _{eq} = 0.022 \pm 0.001$. The value is in excellent agreement with the relationship between the surface shear rate and the rotational period derived by \citet{kovari2017}, who made a statistic study on the differential rotation and found the different laws of rotation-surface shear for single and binary stars. The differential rotation of binary stars is suppressed by the tidal force \citep{petit2004}.

We have clearly detected the weak absorption lines of the third component of SZ Psc from the high-SNR LSD profiles. From the measured radial velocity curves of the inner binary system and the outer tertiary star, we have derived an elliptical orbit with a period of 1530 days and a mass of 0.75 M$\odot$ for the tertiary component. The period of the outer orbit is apparently larger than 1280 days derived from 2004 and 2006 datasets, which was based on the assumption of a circular orbit \citep{xiang2016}, but is very consistent to the period estimated by \citet{eaton2007}. They inferred possible periods of 1143 or 1530 days for the third star and put an upper limit of 1600 days, corresponding to a reasonable total mass of the triple system. Currently, we have a relatively complete phase coverage for the wide orbit of SZ~Psc, but the number of data is still limited (Figure \ref{fig:rv}). We need more observations to refine the orbital solution, especially those around the periapsis passage. The outer orbit of a triple system can be very complex and highly eccentric \citep{mahy2018,tokovinin2023}.

Our monitoring on SZ Psc is still ongoing. With more observations, we will hopefully reveal more details on the long-term spot activity of the active K subgiant and uncover the property of the tertiary component of SZ Psc in the future.

\begin{acknowledgments}
We are very grateful to the anonymous referee for the valuable comments and suggestions. This study is supported by the National Natural Science Foundation of China under grants Nos.10373023, 10773027, U1531121, 11603068, 11903074, and 12373039. We acknowledge the science research grant from the China Manned Space Project. The present study is also supported by the Yunnan Fundamental Research Projects (grant Nos. 202201AT070186 and 202305AS350009), the Yunnan Revitalization Talent Support Program (Young Talent Project), and International Centre of Supernovae, Yunnan Key Laboratory (No. 202302AN360001). We acknowledge the support of the staff of the  Xinglong 2.16m telescope. This work is partially supported by the Open Project Program of the Key Laboratory of Optical Astronomy, National Astronomical Observatories, Chinese Academy of Sciences. This work has made use of the VALD database, operated at Uppsala University, the Institute of Astronomy RAS in Moscow, and the University of Vienna.
\end{acknowledgments}

\bibliographystyle{aasjournal}
\bibliography{szpsc2014}

\appendix

\section{Fits to the observed LSD profiles}

\begin{figure}
\centering
\includegraphics[width=0.3\textwidth]{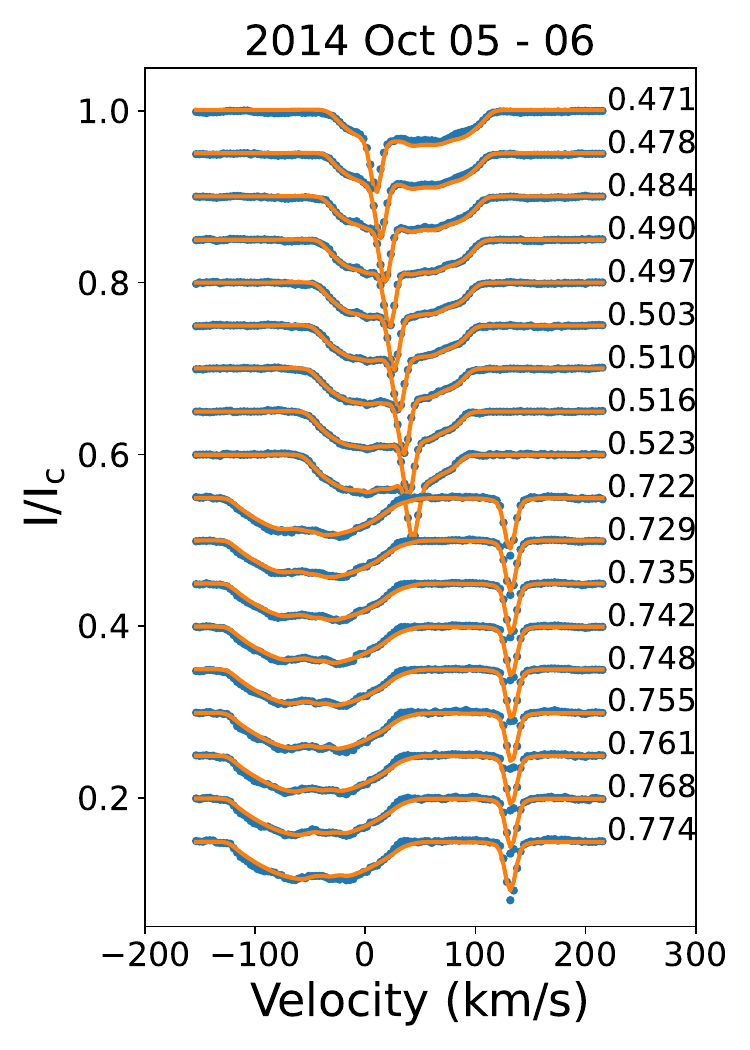}
\includegraphics[width=0.3\textwidth]{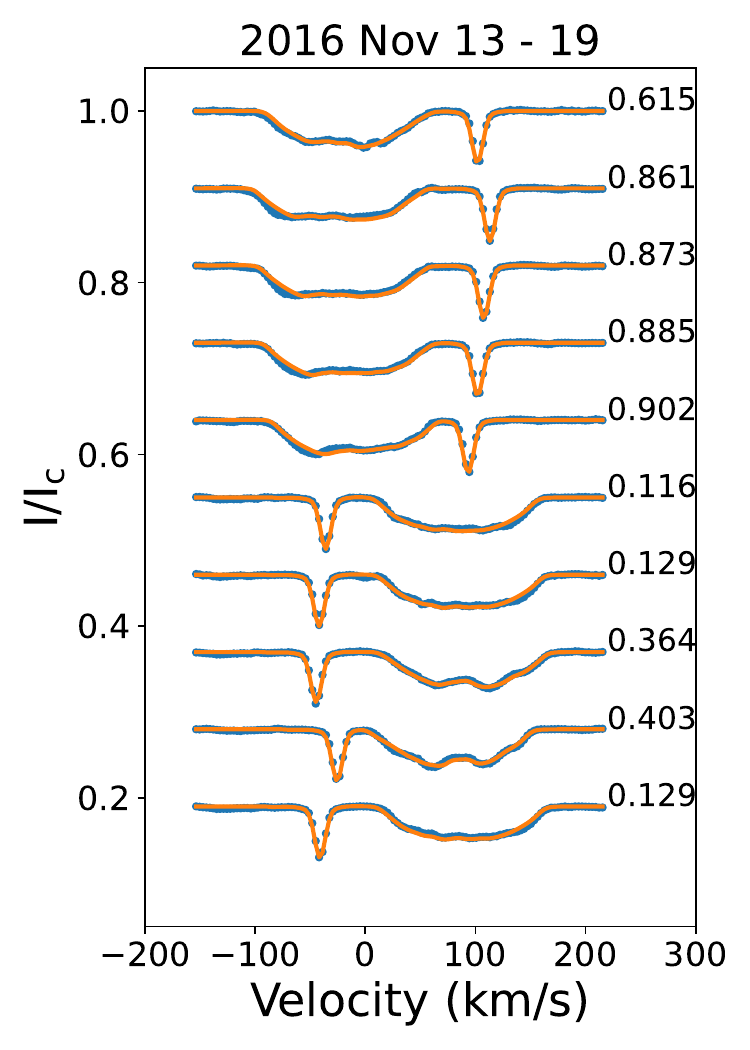}
\includegraphics[width=0.3\textwidth]{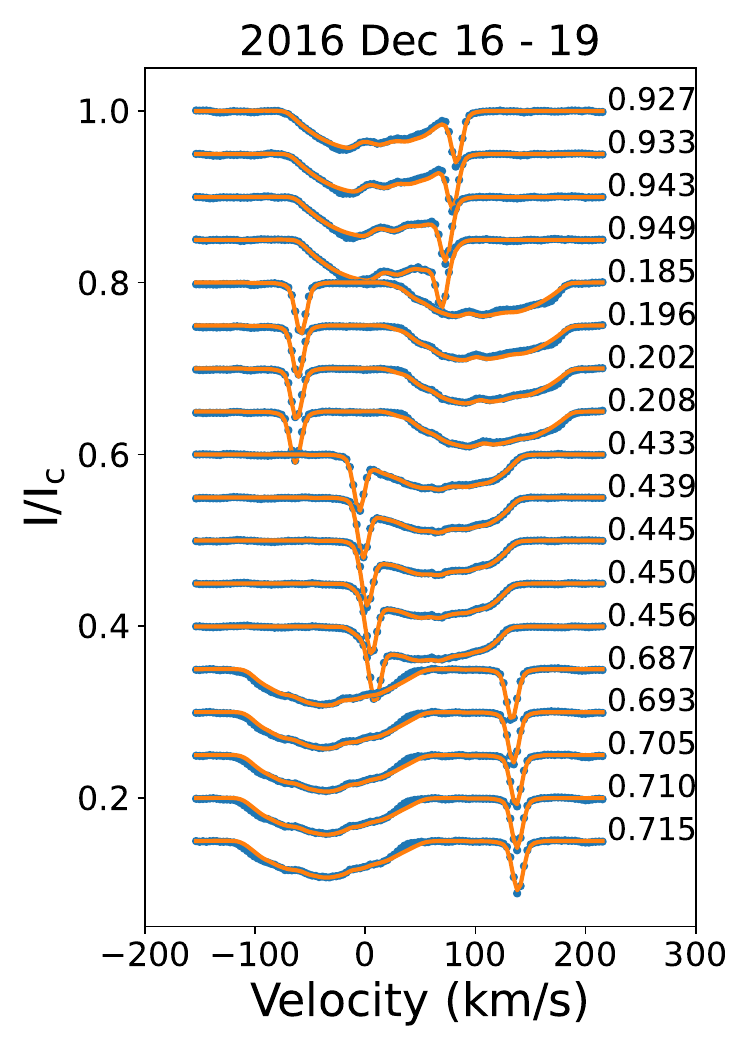}
\includegraphics[width=0.9\textwidth]{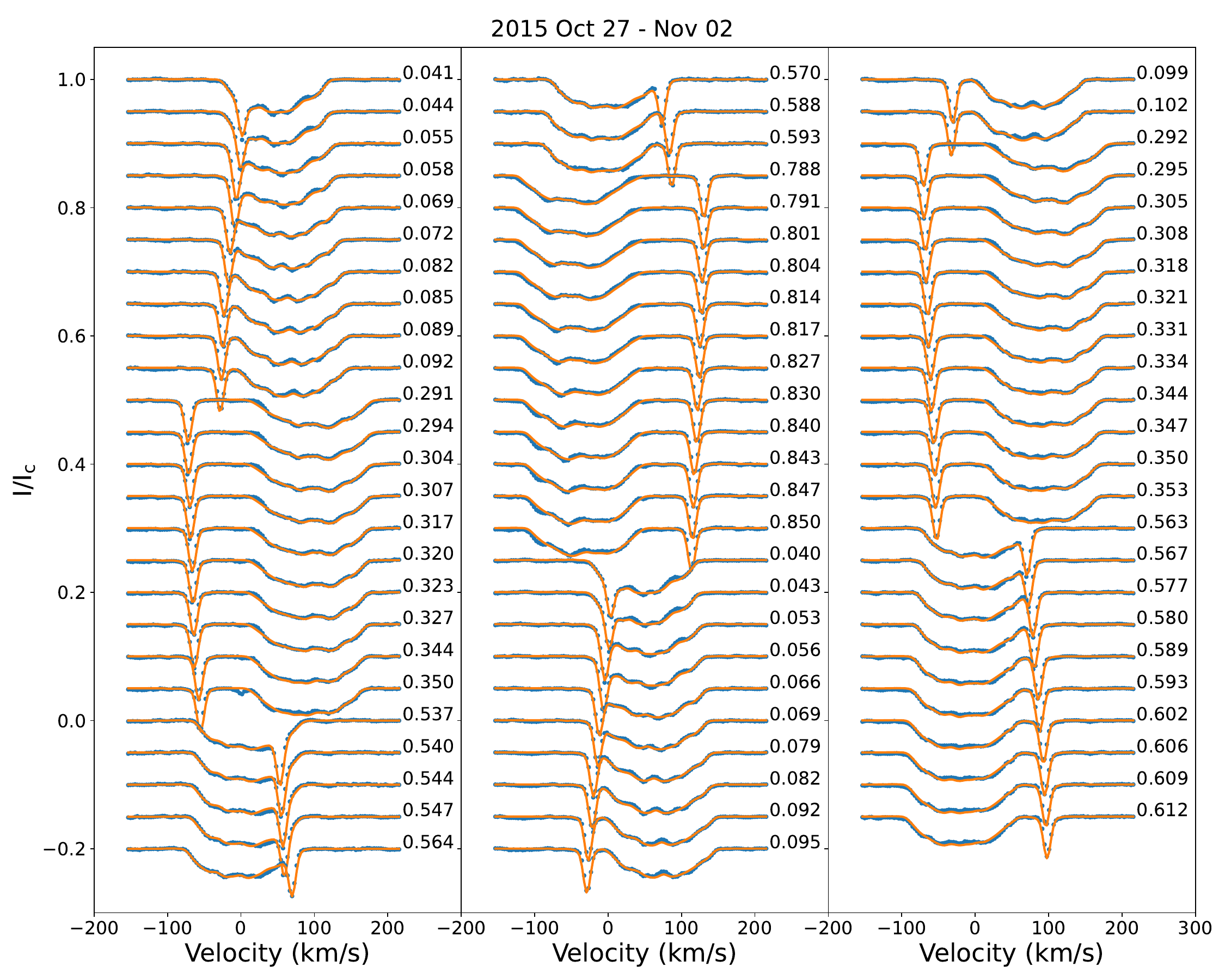}
\caption{The maximum entropy regularized fits to the observed LSD profiles of SZ~Psc for 2014, 2015, 2016 datasets.}
\label{fig:fits1}
\end{figure}

\begin{figure}
\centering
\includegraphics[width=0.3\textwidth]{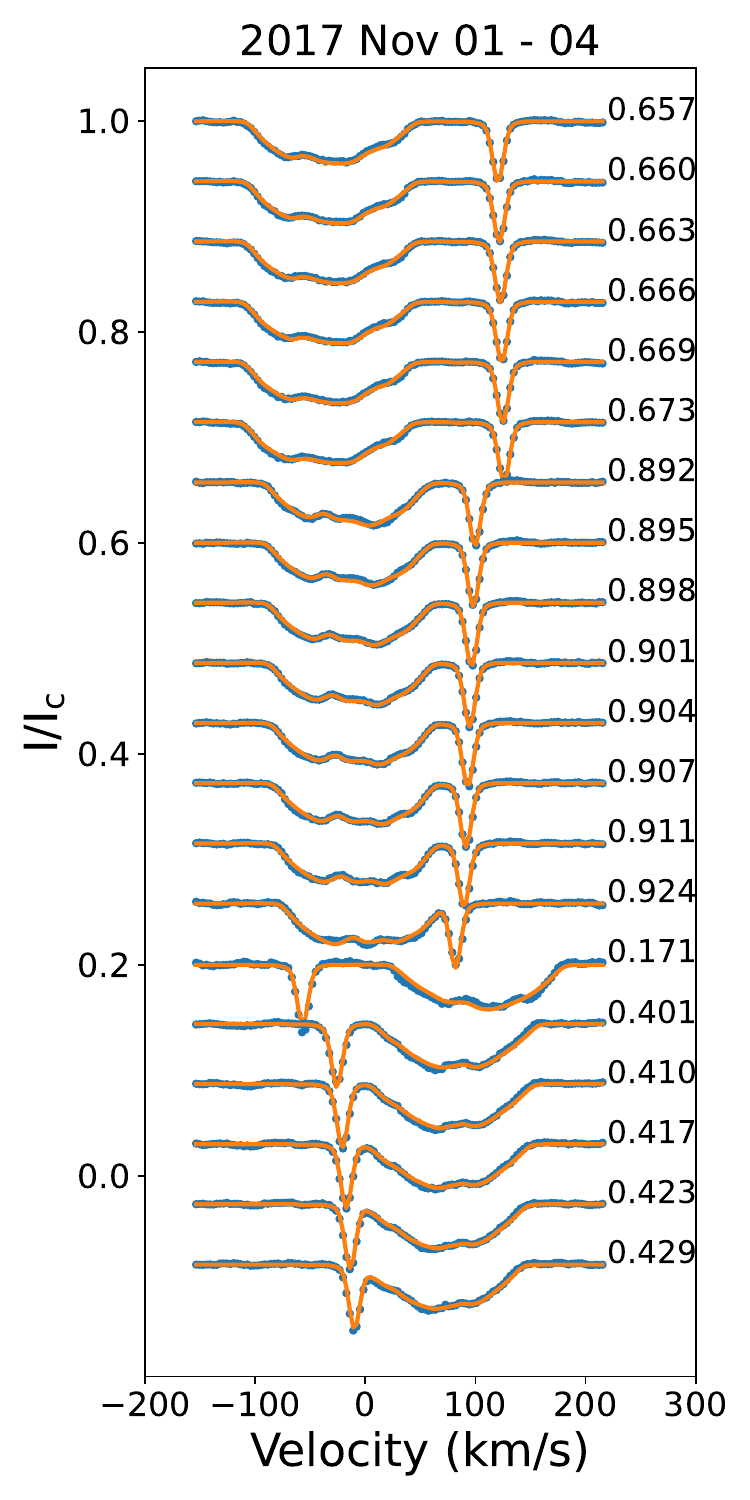}
\includegraphics[width=0.6\textwidth]{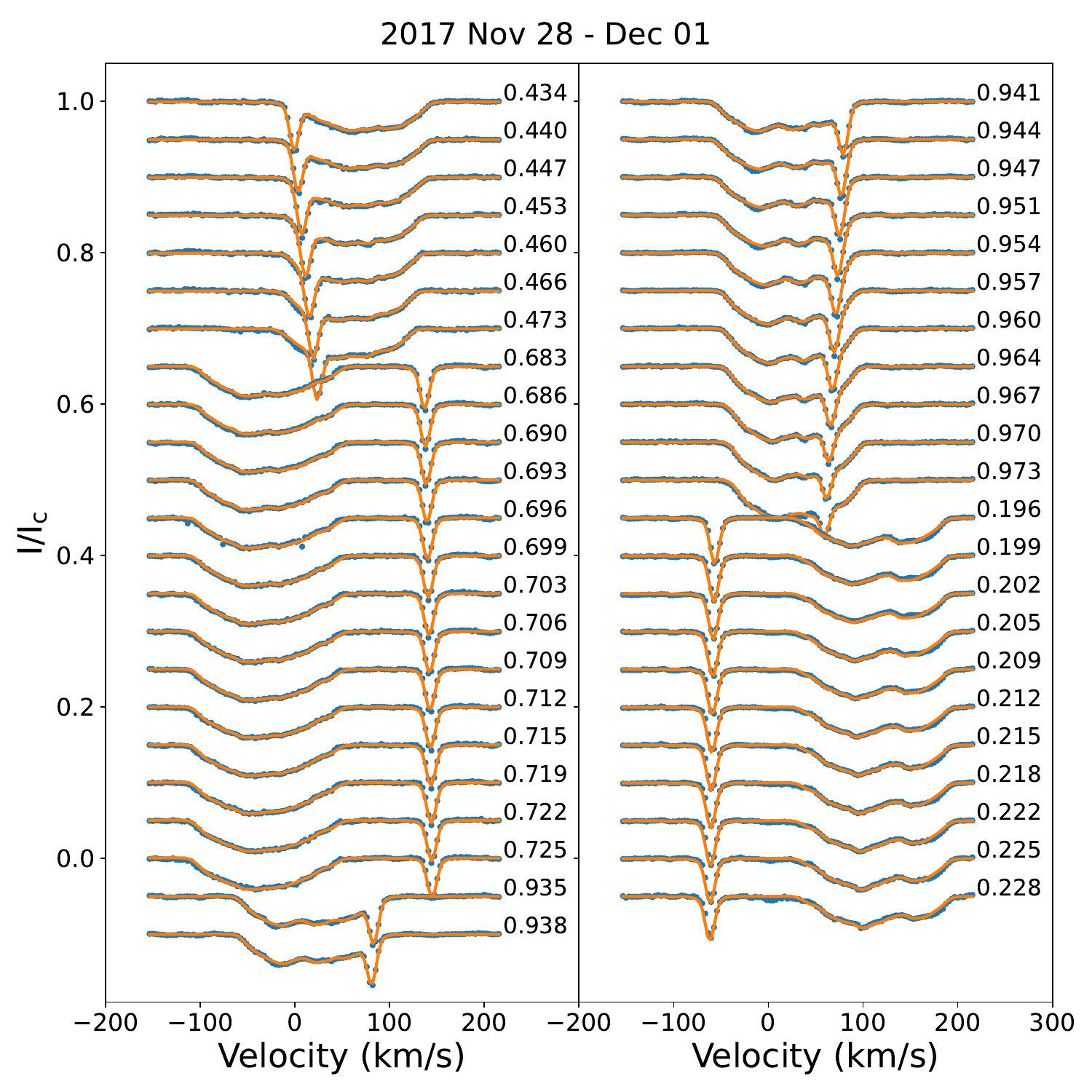}e
\includegraphics[width=0.8\textwidth]{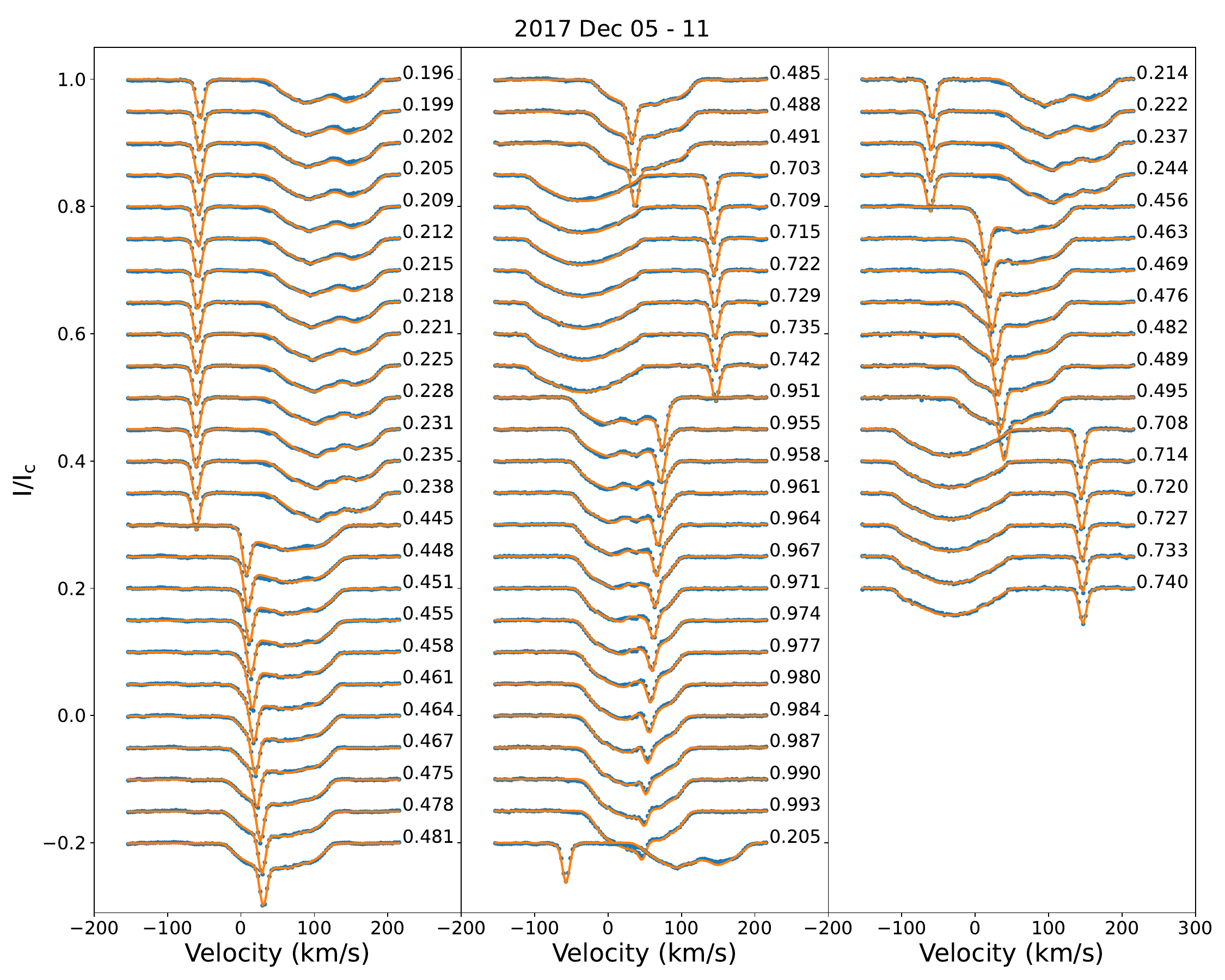}
\caption{Same as Figure \ref{fig:fits1}, but for 2017 datasets.}
\label{fig:fits2}
\end{figure}

\begin{figure}
\centering
\includegraphics[width=0.3\textwidth]{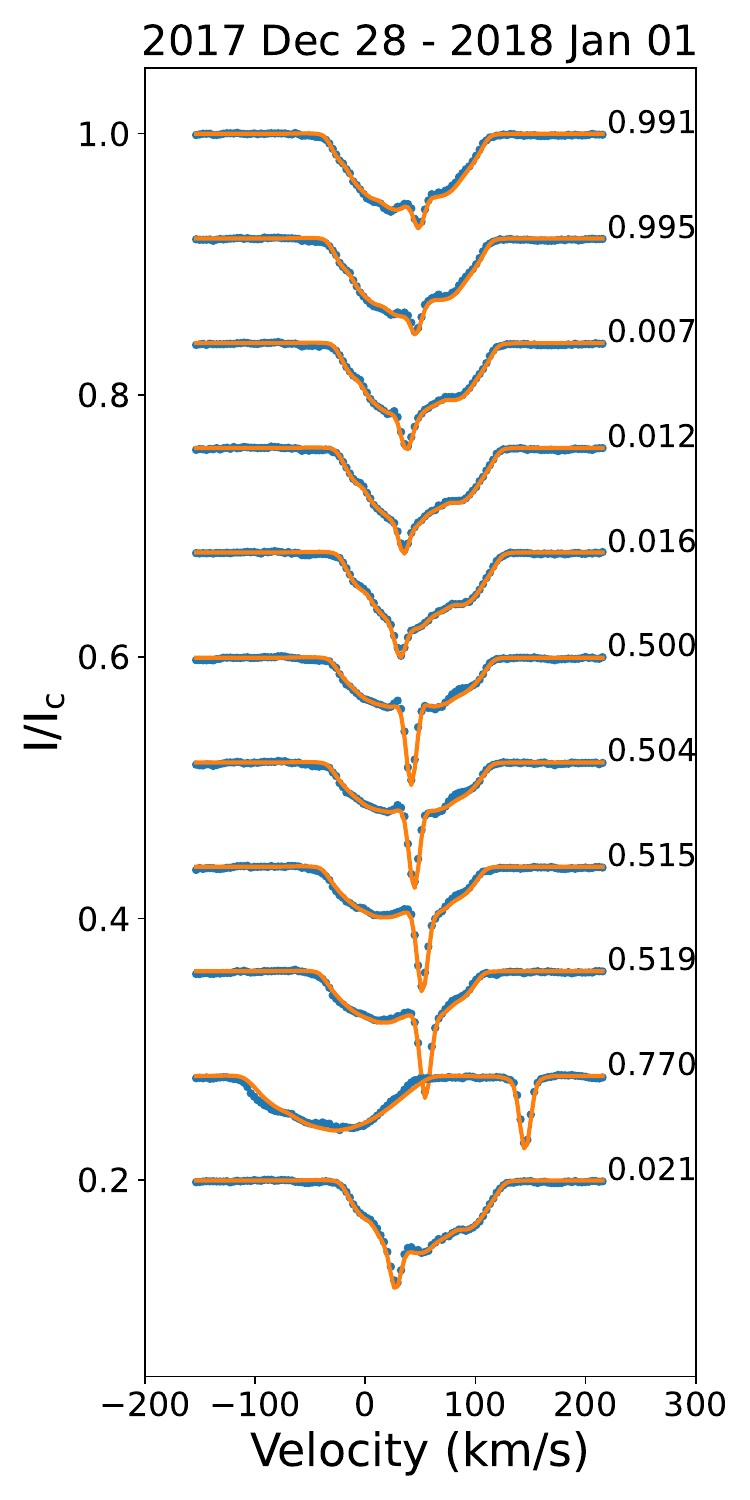}
\includegraphics[width=0.6\textwidth]{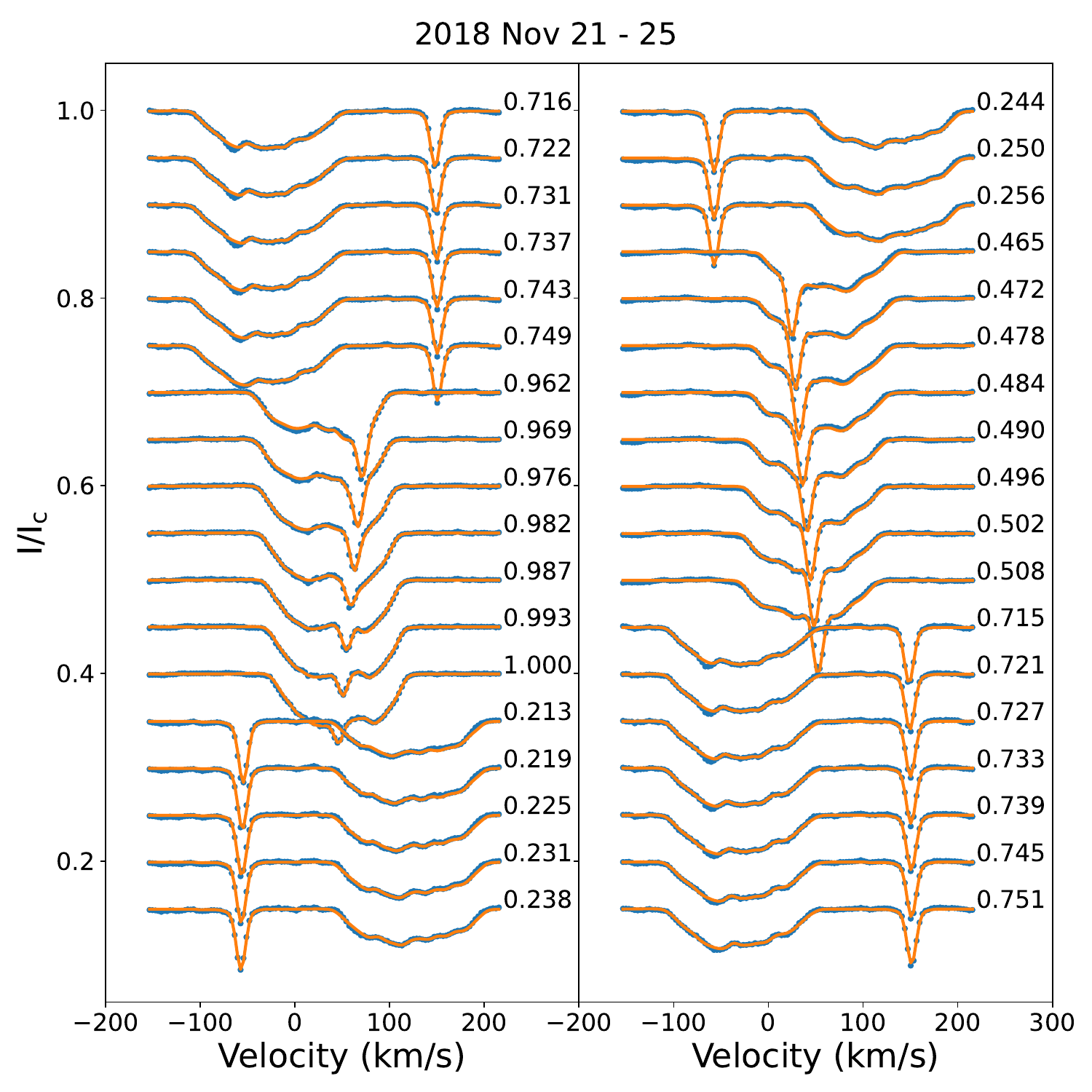}
\caption{Same as Figure \ref{fig:fits1}, but for 2017-2018 datasets.}
\label{fig:fits3}
\end{figure}

In Figure \ref{fig:fits1}--\ref{fig:fits3}, we show the maximum entropy regularized fits derived by DoTS as well as the observed LSD profiles.

\end{CJK*}
\end{document}